\documentclass[manuscript]{acmart}

\usepackage{subcaption}
\usepackage{makecell}
\usepackage{xcolor,soul}
\usepackage{fancybox}

\usepackage{listings}
\acmConference[Conference acronym 'XX]{Make sure to enter the correct
  conference title from your rights confirmation emai}{June 03--05,
  2018}{Woodstock, NY}
\acmISBN{978-1-4503-XXXX-X/18/06}

\begin{document}

\title{LLMCode: Evaluating and Enhancing Researcher-AI Alignment in Qualitative Analysis}


\author{Joel Oksanen}
\email{joel.oksanen@aalto.fi}
\author{Andrés Lucero}
\email{andres.lucero@aalto.fi}
\author{Perttu Hämäläinen}
\email{perttu.hamalainen@aalto.fi}
\affiliation{%
  \institution{Aalto University}
  \city{Espoo}
  \country{Finland}
}

\renewcommand{\shortauthors}{Oksanen et al.}

\begin{abstract}

The use of large language models (LLMs) in qualitative analysis offers enhanced efficiency but raises questions about their alignment with the contextual nature of research for design (RfD). This research examines the trustworthiness of LLM-driven design insights, using qualitative coding as a case study to explore the interpretive processes central to RfD. We introduce LLMCode, an open-source tool integrating two metrics—Intersection over Union (IoU) and Modified Hausdorff Distance—to assess the alignment between human and LLM-generated insights. Across two studies involving 26 designers, we find that while the model performs well with deductive coding, its ability to emulate a designer's deeper interpretive lens over the data is limited, emphasising the importance of human-AI collaboration. Our results highlight a reciprocal dynamic where users refine LLM outputs and adapt their own perspectives based on the model's suggestions. These findings underscore the importance of fostering appropriate reliance on LLMs by designing tools that preserve interpretive depth while facilitating intuitive collaboration between designers and AI.


\end{abstract}




\begin{teaserfigure}
    \centering
    \includegraphics[width=\textwidth]{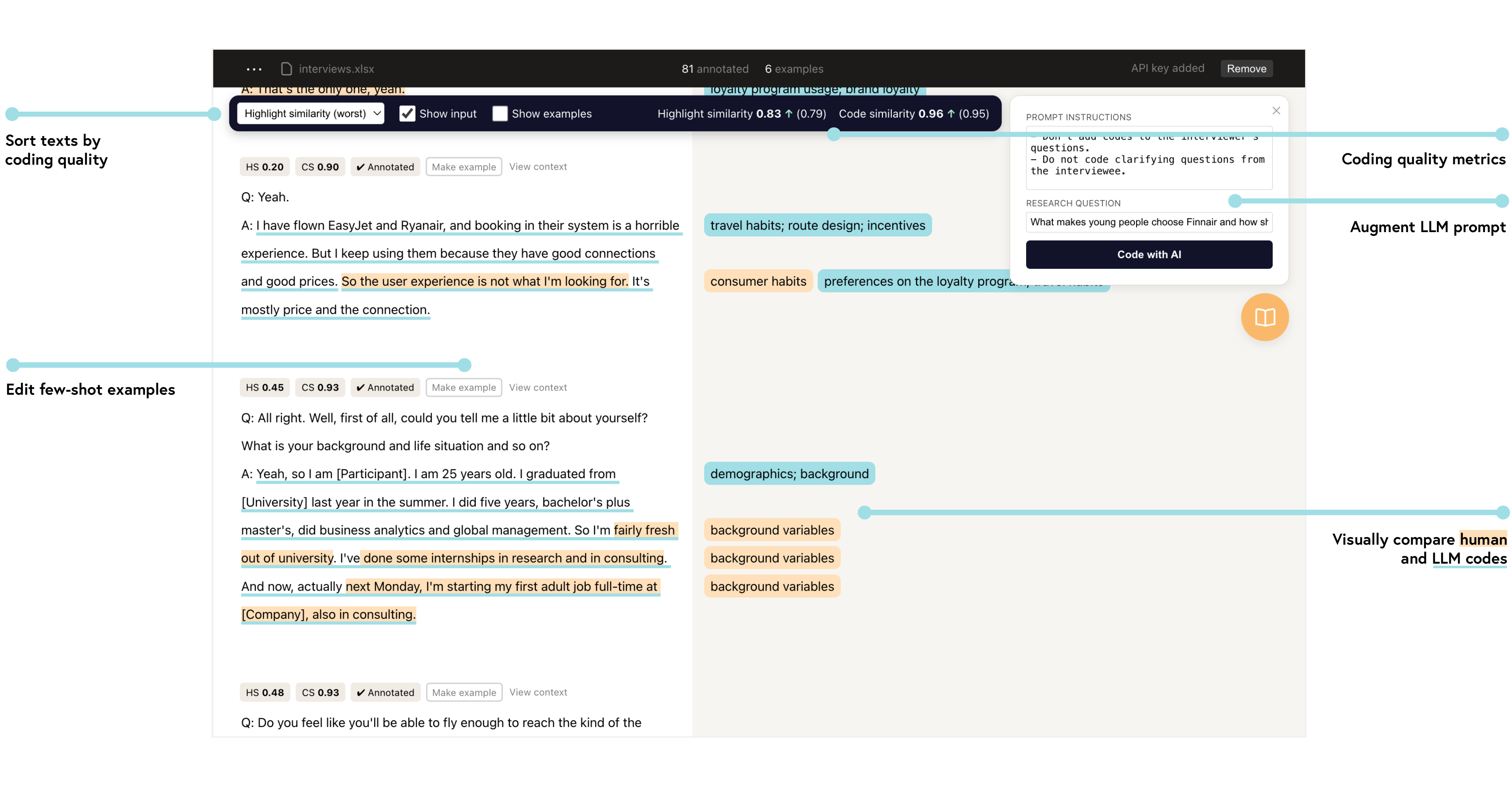}
    \caption{LLMCode interface for researcher-aligned LLM-assisted qualitative coding. The interface enables users to compare the model's output to their own coding, identify misalignments between the human and LLM annotations, and correct these through iterative refinement. To aid this process, texts can be sorted using two coding quality metrics, helping users identify areas for improvement. Users can then adjust the prompt instructions and coding examples provided to the model to enhance alignment.}
    \Description[]{A screen capture showing the LLMCode interface for researcher-aligned LLM-assisted qualitative coding. The interface shows a large open area with texts that have been coded by the human in yellow and the AI system in blue. At the top, there are metrics and other functionality to help the user navigate the interface. The interface enables users to compare the model's output, identify misalignments between human and LLM annotations, and correct these through iterative refinement. To aid this process, texts can be sorted using two coding quality metrics, helping users identify areas for improvement. Users can then adjust the prompt instructions and coding examples provided to the model to enhance alignment.}
    \label{fig:interface}
\end{teaserfigure}

\maketitle

\section{Introduction}

Designers’ work often builds upon insights derived from qualitative analysis, where human-generated, often textual, data, such as interview transcripts or online discussions, are systematically examined to identify patterns, thoughts, and behaviours. There has been an ongoing shift towards this form of evidence-based design\footnote{While specialised research-focused design roles such as "user researcher" exist, many design roles fall under the concept of a "practitioner-researcher" \cite{Gray:04}, covering a range of responsibilities, one of which being research. For this reason, we use the terms "designer" and "researcher" interchangeably throughout the paper.} across design disciplines \cite{Muratovski:22}, wherein insights extracted from human-centred data are used to inform and justify the design choices for new artefacts, such as products or services.

In parallel with this increasing research orientation, both academia and industry have actively sought ways to offload the more laborious aspects of qualitative analysis to automated systems, a pursuit that is often motivated by the intention of increasing the quantity of data that can be feasibly analysed \cite{Ashwin:23} and freeing up time for more creative tasks such as ideation \cite{Dunvin:24}. The recent emergence of large language models (LLMs) has accelerated this trend, as they exhibit an unprecedented ability to process, summarise, and generate human-like text, thereby offering a promising route for automating various stages of qualitative analysis. In fact, several tools and workflows for LLM-driven qualitative coding have been proposed in recent years, and numerous commercial platforms, such as Atlas.ti\footnote{https://atlasti.com} and Dovetail\footnote{https://dovetail.com}, have begun advertising AI features that claim to transform complex qualitative data into “instant insights” \cite{Dovetail:24}.

Few studies, however, have addressed what it means to evaluate the quality of these LLM-generated insights. Prior work in human factors has shown that the success of automation hinges on humans’ appropriate reliance on it \cite{lee:04}, making researcher-AI trust—a precursor to reliance—an important factor in the effective adoption of AI within research for design (RfD) \cite{Oksanen:24}. Yet, because evaluation is a key method for establishing warranted human-AI trust \cite{jacovi:21}, the lack of tailored assessment methods for this context may leave practitioners unsure whether they can rely on AI-driven insights. There is already evidence of such a phenomenon: while contemporary Computer-Assisted Qualitative Data Analysis Software (CAQDAS) have long offered increasingly sophisticated AI features, it has been shown that many researchers only engage with them at a superficial level—often treating the software as little more than an “electronic filing cabinet” \cite{Marathe:18} for their manual analysis practices. The significance of this research gap is compounded by the inherent opaqueness of many qualitative research approaches, which may, if left unaddressed, lead to situations where low-quality AI-generated research displaces human work, particularly in settings where costs are the primary driver of decision-making.

In many qualitative methods employed by design practitioners, such as thematic analysis or affinity diagramming \cite{Lucero15}, the designer’s introspection and personal interpretation of the data play a central—and necessary—role in shaping the outcome \cite{Braun:19, Schon:83}. This further complicates the question of appropriate reliance when introducing AI-driven tools: on one hand, the quest for improved scale and efficiency incentivises collaborative systems in which the designer cedes some autonomy to a machine partner; 
on the other hand, these AI collaborators lack direct access to the designer’s tacit knowledge and, by design, may risk homogenising perspectives rather than supporting the unique, reflexive stance that RfD thrives on \cite{Bender:21}. Given that there is no single source of truth in such interpretive approaches, it follows that one measure of trustworthiness for an AI system may investigate how closely it can align with the perspectives of an individual designer who is deeply immersed in a given context.

Building on these observations, our research addresses the following question: 
\begin{itemize}
    \item \textbf{RQ 1:} How can we measure the alignment between AI-generated insights and a designer’s interpretive insights?
\end{itemize}
By developing tailored evaluation metrics for this context, we aim to ensure appropriate reliance on AI-assisted research tools. These metrics not only provide a foundation for evaluating trustworthiness, but also allow us to explore how such tools could preserve the diversity of perspectives inherent in design and avoid diminishing the depth of designers' insights. To further investigate these dynamics, we pose two additional research questions:
\begin{itemize}
    \item \textbf{RQ 2:} To what extent can an LLM research tool emulate an individual designer’s perspective on data?
    \item \textbf{RQ 3:} How and to what extent does LLM assistance shape the research insights generated in a design context?
\end{itemize}

Whilst acknowledging the diverse research practices and methods of designers, in this study we focus on the specific task of qualitative coding in order to investigate how insights are shaped with and without AI assistance in a structured, transparent, and quantifiable setting. Coding is a common practice across a wide range of qualitative research domains, which entails applying labels to segments of qualitative data to systematically capture insightful patterns, concepts, and ideas \cite{Miles:14}.

To this end, we introduce a novel open source qualitative coding tool, LLMCode, that integrates two new metrics for the evaluation of LLM-driven qualitative coding, inspired by established machine learning approaches: Intersection over Union (IoU) and Modified Hausdorff Distance (MHD). IoU measures how well the LLM identifies salient content by comparing the overlaps in coded segments between the model and a human baseline, thereby indicating how closely the model’s focus aligns with that of a human researcher. Meanwhile, MHD, which calculates the embedding distance between the labels generated by the LLM and a human, focuses on whether the insights and themes extracted by the model align with those conceptualised by a human. Together, these metrics help us systematically capture where and how the LLM’s lens on the data diverges from the designer’s.

To explore the implications of LLM-driven insights, we employed LLMCode in two empirical studies. The first study involved 19 students engaging with the tool in the context of game design research projects. By comparing participants’ initial manual coding with the same texts coded by LLMCode, we demonstrate that while an LLM may appear to emulate human coding through in-context learning, this capability significantly diminishes when the model is prompted to analyse texts with unfamiliar codes, suggesting that LLMs are limited in their ability to replicate human researchers' inductive reasoning processes. In a subsequent user study involving a smaller, more diverse sample of designers, we examined the reciprocal dynamics of influence between researcher and AI during an iterative human-AI coding process using the LLMCode interface (Figure \ref{fig:interface}). Our findings suggest that while LLMCode’s metrics assist researchers in curating examples to align the model with their perspective, they were equally willing to adapt their own coding practices in response to the model’s outputs, underscoring a bi-directional exchange of insights.

\paragraph{Contribution:} In summary, this work makes two key contributions: (1) introducing a systematic framework for evaluating and aligning LLM-driven coding within the reflexive, constructivist paradigm that underpins research for design (RfD); and (2) applying this framework in two empirical studies, generating novel insights into the dynamics of designer-AI collaboration in research contexts. These findings offer valuable guidance for developers and researchers seeking to create AI tools that preserve the interpretive depth and richness of qualitative analysis, while still offering substantial gains in efficiency and breadth of analysis. Our open-source AI-assisted qualitative coding toolkit LLMCode is published on GitHub: \textit{<link omitted for anonymous review>}.

\section{Background}

\subsection{AI-Assisted Qualitative Research}

Existing work on AI-assisted qualitative research predominantly focuses on academic contexts, particularly on technologies that support qualitative coding—a standardised practice for analysing textual qualitative data. In this process, researchers annotate segments of text, such as field notes or interview transcripts, with short labels, or \textit{codes}, to summarise and capture key insights \cite{Miles:14}. While coding methods vary across researchers and disciplines, this paper centres on inductive coding, where codes are generated during analysis rather than being predetermined \cite{Miles:14}. This approach aligns well with the constructivist nature of RfD, making it particularly suitable for studying insight development in design practice.

\subsubsection{AI-Assisted Qualitative Coding}

Numerous AI-assisted workflows have been proposed for coding, offering varying levels of automation. Early contributions to the inductive domain include Scholastic by \citet{Hong:22}, an interactive system designed to support inductive coding by employing topic modelling to generate code suggestions. Similarly, PaTAT by \citet{Gebreegziabher:23} introduces a novel interface that enables researchers to iteratively refine AI-suggested patterns.

More recently, a growing body of research has explored the potential of LLM-based systems for automated coding, offering capabilities that go beyond traditional topic modeling techniques. Unlike topic modeling, which typically represents a category of texts through selected keywords, LLMs can be provided with a sample of texts and prompted to generate a human-readable description of the category. Several case studies have examined the use of out-of-the-box LLMs, such as OpenAI's GPT-3 \cite{Brown:20} and ChatGPT \cite{OpenAI:22}, for fully automated coding across a diverse range of qualitative research projects, including media analysis \cite{Dunvin:24, Hamalainen:23}, phenomenological research \cite{Hamilton:23}, and the social sciences \cite{Xiao:23}. Argument2Code \cite{Zhao:24} exemplifies a more sophisticated automated system, focusing on inductive codebook generation through a multi-step process employing chain-of-thought prompting.

Another line of research has explored the development of human-LLM collaboration frameworks for coding. \citet{Hamalainen:23} and \citet{Dai:23} demonstrated that LLM coding is possible using in-context learning \cite{Brown:20}, by including human-annotated \textit{few-shot} examples of coded texts in the prompt. This process can be further refined through an iterative dialogue between the researcher and the system \cite{Dai:23}. \citet{Lopez:24} investigate methods to enhance the transparency of human-LLM coding through visualisations and manual tracing of codebook development. They found that 39.28\% of their final codebook comprised codes influenced by LLM suggestions during the coding process. Similarly, \citet{Sinha:24} reflect on their use of GPT-4 to support grounded theory analysis, observing that the model facilitated the discovery of new codes. 


\subsubsection{Evaluating Outputs and Researcher Perspectives}

Most of the aforementioned studies evaluate the LLM-generated outputs either qualitatively \cite{Hong:22, Gebreegziabher:23, Dunvin:24, Hamilton:23, Lopez:24, Sinha:24} or using Cohen's Kappa \cite{Xiao:23, Dai:23, Dunvin:24}, a widely employed metric for measuring inter-annotator agreement in collaborative coding. However, Cohen's Kappa is only applicable in deductive coding, where the annotators use a shared codebook. Additionally, \citet{Dai:23} and \citet{Zhao:24} proposed methods using cosine similarity between word embeddings to measure the semantic similarity of human- and LLM-generated codebooks. Building on this, our work introduces an embedding-based method to compare code annotations for individual inductively coded texts.

Need-finding studies on AI-assisted qualitative coding by \citet{Jiang:21} and \citet{Feuston:21} have examined researchers’ attitudes towards human-AI collaboration in qualitative research, motivated by the growing disparity between advancing coding tools and comparatively outdated industry practices. While these studies highlight tedious sub-processes within coding that could be streamlined through AI methods, the authors caution against the complete automation of research, stressing the importance of maintaining user agency and supporting serendipitous insight discovery. Specifically, they argue that AI should act as a supportive assistant rather than an “eager” collaborator \cite{Feuston:21}, whose unsolicited suggestions might undermine the researcher’s central role.


Several studies have taken a critical perspective on the use of LLMs for qualitative analysis. \citet{Ashwin:23} employ ChatGPT (gpt-3.5-turbo) and Llama 2 (13b) \cite{Touvron:23} in coding semantically complex semi-structured interview data, demonstrating that the outputs often exhibit systematic bias, potentially leading to misleading interpretations in subsequent analysis. They emphasise the necessity of human expert annotations to validate LLM outputs and suggest that LLMs are best suited for extending traditional qualitative analysis to larger corpora—provided the analysis begins on a smaller scale with the more nuanced manual work of human researchers. Similarly, while \citet{Sinha:24} found LLMs useful for identifying gaps in their own grounded theory analysis, they caution that relying on such tools could discourage the deep immersion in data required by grounded theory, potentially leading to superficial interpretations. Rather than fully depending on LLMs for coding, the authors recommend researchers independently code their data before comparing it with LLM-generated outputs.

\subsection{Trust in Automation}

The trust an operator places in an automated system is widely recognised as a significant factor contributing to appropriate reliance and the success of automation within organisations \cite{lee:04}. Research shows that people respond to technology socially \cite{lee:04}, 
which means that trust may hinge on factors akin to interpersonal trust, such as perceived ability, integrity, and benevolence \cite{mayer:95}. However, ideally, trust in automation should align with the system’s actual capabilities—its \textit{trustworthiness}—such that the operator does not place too little (i.e., distrust) or too much (i.e., overtrust) trust upon the system, which may lead to under- or overreliance, respectively \cite{lee:04}.

In the context of human-AI trust, \citet{jacovi:21} define \textit{warranted trust} as trust that arises from an AI system’s trustworthiness. According to their framework, warranted trust can be established through either \textit{intrinsic} or \textit{extrinsic} factors. Intrinsic trust is closely tied to explainability and is developed when the AI’s observable decision-making process aligns with the human’s expectations of how the process should function. In contrast, extrinsic trust is based on an evaluation of the AI’s behaviour or outputs. This encompasses not only the model’s performance but also confidence in the evaluation methods used to assess the system.


Building on the work of \citet{jacovi:21}, \citet{Oksanen:24} highlights the difficulty of objective insight evaluation as a barrier to the development of extrinsic trust in the context of AI-assisted RfD. In the following section, we build a case for evaluating AI-driven design insights based on their similarity to an \textit{individual} designer's insights. This approach positions AI-assisted design research as an inherently collaborative process, where the AI system and the designer work together, with the designer’s choices serving as a baseline for comparison.

\subsection{Research for Design}
\label{sec:design}

In this paper, we focus on \textit{research for design}, defined as the process of need-finding and exploration that takes place within many practical design project contexts \cite{Clemente:17}. The necessity of research within design projects stems from the increasing complexity of design briefs, which require designers to address not only form and style but also “the broader context in which their designs will be used” \cite{Muratovski:22}. This research is frequently human-centred and qualitative in nature, aiming to foster empathy for the end users that may "inform and inspire" the artefact that is being created \cite{Sanders:08}.

At the same time, unlike many forms of research conducted within traditional science, RfD does not aim for absolute objectivity by grounding every insight in verifiable evidence. On the contrary, designers are regarded as “expert subjects” valued for their “distinctive ways of seeing and doing their work” \cite{Bardzell:16}. In other words, RfD is inherently reflexive, with each designer bringing a unique, situated perspective to a project; their background, experiences, and context all shape how they interpret problems and envision solutions \cite{Gray:04}.

This individuality and reflexivity underpins the diversity of thought that design thrives on. Interdisciplinary collaboration is considered one central aspect of design thinking: as diverging perspectives converge in a collaborative setting, it may result in innovative ideas that break from the mainstream \cite{Cross:11, Micheli:19}. As proposed by \citet{Krippendorff:05}: "Design is fundamentally concerned with innovation, with making changes happen, and designers are especially challenged by common beliefs in what cannot be done." Accordingly, RfD is not only concerned with how this \textit{are}, but also about uncovering and inspiring novel ideas for how things \textit{could be} \cite{Dunne:13}. 

These prevalent theorisations of design practice introduce a tension when contrasted with the emerging use of LLM-driven systems in research. LLMs, by virtue of their training, have been described as “stochastic parrots” \cite{Bender:21}, in that they are trained on vast corpora of text and generate outputs reflecting aggregate surface patterns already present in that data. Such systems are—by design—"average" thinkers, which is in direct juxtaposition to the aforementioned objectives of design to envision novel futures that have not yet transpired \cite{Lucero18}, by the way of individual perspectives. If designers allow LLMs to strongly shape or homogenise their thought processes, it could flatten the diversity of perspectives and reduce the likelihood of genuinely novel insights. This effect is further exacerbated if the same AI system is used by many designers, nudging everyone toward similar conclusions or ideas.

Due to the subjective and contextual nature of RfD, design insights can be challenging to evaluate objectively. This complexity, combined with the potentially superficial yet highly convincing outputs generated by LLMs \cite{Bender:21}, raises questions about the risk of overtrust—and consequently, overreliance—on LLM-driven insights in this context. While further investigation is needed, there is a potential concern that LLM systems could unduly influence designers’ situated interpretation of data. Conversely, systems that focus on \textit{emulating an individual designer’s viewpoint}, rather than asserting their own generic perspective, may offer promising opportunities. Such systems could potentially scale research to much larger corpora while enabling each designer to maintain their personal reflexive stance.

\section{LLMCode}

LLMCode is an open-source toolkit for qualitative coding with LLMs. It supports a workflow based on initial manual coding and in-context learning, similar to that proposed by \citet{Hamalainen:23} and \citet{Dai:23}. This approach requires—in addition to the raw input texts to be coded by the model—a smaller example set that has already been manually coded by a human researcher. LLMCode is differentiated from other qualitative coding tools by its novel integration of two quality metrics—IoU and MHD—to evaluate and improve the alignment between the human user's and the LLM's coding.

The toolkit was developed through an iterative design process, with the aim of improving the trustworthiness of LLM-driven qualitative coding. In the first of our two studies using the system (Section \ref{sec:study1}), participants interacted with the LLMCode functions in a Jupyter Notebook environment \cite{Kluyver:16}, where they were able to call upon LLMCode's functions through Python code and produce tables and visualisations of its output. Based on this study, we identified that in order to fully take advantage of the system's capabilities, users would require a custom user interface. Using our findings from the first study, such an interface was subsequently developed and evaluated in the second study (Section \ref{sec:study2}).

LLMCode includes functions for both inductive and deductive coding. In the studies reported in this paper, participants used the toolkit's \texttt{code\_inductively\_with\_code\_consistency} function for automated inductive coding. The system prompts an LLM to annotate each text using markdown notation, where coded segments (\textit{highlights}) are bolded by surrounding them with double asterisks \texttt{**}, immediately followed by a list of code labels enclosed in superscript \texttt{<sup>} tags. This structured output format allows the model to place several highlights with distinct codes across a single text, where most existing work on LLM-assisted coding has focused on the assignment of codes to entire texts. The function processes texts sequentially, and each newly generated code is added to a \textit{codebook} which is included in subsequent prompts. Similar to manual inductive coding, the codebook is initially empty and expands as new codes emerge during the process. For datasets where context is crucial, the system accommodates contextual information within the prompt, such as preceding messages in an online discussion thread. An example of the full prompt structure is shown in Appendix \ref{appendix:prompt}.

The model is instructed to reproduce each input text verbatim, adding only the code annotations, in order to retain the full meaning of the original data. However, the input data may occasionally contain spelling variations or other irregularities that the model was observed to sometimes “correct” despite explicit instructions not to alter the text. To ensure that the model does not hallucinate or change the data beyond the correction of minor spelling errors, LLMCode's functions check each model output for discrepancies and attempt to correct these through reconstructing the annotations onto the original text using approximate string matching.

\subsection{Coding Quality Metrics}
\label{sec:metrics}
LLMCode utilises Intersection over Union (IoU) and Modified Hausdorff Distance (MHD) to quantify human-LLM alignment focusing on both: (1) what text is highlighted; and (2): how these highlights are coded. Table \ref{table:human_vs_llm} illustrates the two metrics with interview data examples.

\subsubsection{Intersection over Union} IoU, also known as Jaccard Index~\cite{real1996probabilistic}, assesses the alignment of LLM and human highlights, providing insights into how well the LLM identifies key content and how closely the model's focus matches that of a human researcher. Specifically, IoU expresses the overlap of human and LLM highlights with a score that is between zero (no overlap) and one (exactly the same text highlighted), and accounts for differences in both the scope and precision of the LLM’s highlights relative to human annotations. 
IoU is a common success metric in other areas of machine learning, e.g., for measuring the overlap of object detection bounding boxes~\cite{rezatofighi2019generalized} or the quality of extractive text summarization~\cite{verma2017extractive}. LLMCode implements IoU on the level of individual characters. 

\subsubsection{Modified Hausdorff Distance} We utilise the MHD measure \cite{dubuisson1994modified} to evaluate the alignment between codes generated by an LLM and those created by a human researcher. MHD is common way to calculate the similarity between two point clouds in computer vision. Here, we use an LLM-based embedding model to map each code to a point in the embedding space, and construct LLM and human point clouds from the codes applied to each coded text. MHD between such point clouds captures relatedness in meaning, even when codes differ in ordering, count, wording, or emphasis. 

The interpretation of MHD depends on the embeddings used, which are shaped by the specific embedding model and its training data. While the values may reflect model-specific biases and are not absolute, they provide a valuable lens for examining alignment by highlighting nuanced differences in how codes are represented and associated. Together with IoU, MHD offers a more comprehensive evaluation by combining surface-level agreement with semantic alignment, enabling a deeper understanding of human-AI coding correspondence. 

\subsection{Interface}
\label{sec:interface}

A user interface (Figure \ref{fig:interface}) was developed for LLMCode based on findings from the first study. The interface is designed to incorporate the aforementioned metrics into an iterative process, enabling researchers to align the model with their perspective on the data by editing the coding instructions given in the LLM prompt and selecting a representative set of few-shot examples. Initially, the tool operates similarly to commercial qualitative coding tools: users manually highlight text segments on the left side using a cursor and enter the corresponding codes into a dedicated space on the right. After annotating a sufficient number of texts (for a discussion on what can be considered as "sufficient", see Section \ref{sec:sufficient}), users select a subset of these annotations to serve as the model's few-shot examples. The rest of the manually annotated texts are used to calculate the quantitative alignment metrics. 

The LLM-generated annotations are displayed alongside the human annotations in a distinct colour, allowing users to visually compare and identify where the model's interpretation diverges from their own. The quality metrics are shown both on average and for each coded text. To enable quickly identifying and inspecting problem cases, the texts can be sorted based on the metrics. These insights guide users in iteratively refining the examples and prompts provided to the model, with progress tracked through changes in the metric averages.

Once the user is satisfied with the results both quantitatively and qualitatively, the same coding approach defined by the prompt and examples can be automatically applied to the remaining non-coded data. This process can significantly scale up the amount of data that can be analysed, enabling researchers to handle larger datasets with enhanced efficiency while maintaining oversight of the interpretive process.

\renewcommand{\arraystretch}{1.4}
\sethlcolor{yellow} 
\newcommand{\success}[1]{{#1}} 
\newcommand{\falsepositive}[1]{{#1}} 
\newcommand{\falsenegative}[1]{{#1}} 
\newcommand{\mutualomission}[1]{{#1}} 

\begin{table}[ht]
    \centering
    \small
    \begin{tabular}{|m{0.37\linewidth}|m{0.37\linewidth}|>{\raggedleft\arraybackslash}m{0.08\linewidth}|>{\raggedleft\arraybackslash}m{0.08\linewidth}|}


\hline

\thead{Human annotations} & \thead{LLM annotations} & \thead{IoU} & \thead{Hausdorff \\ distance} \\ 
\hline
\hl{Yeah, I know with Barbie and everything, pink is definitely going up.}\textsuperscript{cultural trends}
& \small
\hl{Yeah, I know with Barbie and everything, pink is definitely going up.}\textsuperscript{improvements}
& 1.00 & 0.15 \\ \hline

\small
I think it starts on basic actually, basic, silver, gold. It's this typical good, better, best. \hl{Maybe they could spice that up a bit}\textsuperscript{improvement suggestions}, make some more remarkable names or different colours. & \small
\hl{I think it starts on basic actually, basic, silver, gold.}\textsuperscript{loyalty program structure} It's this typical good, better, best. \hl{Maybe they could spice that up a bit}\textsuperscript{improvement suggestions}, make some more remarkable names or different colours. & 0.40 & 0.04 \\ \hline

\small

\hl{I travel quite often}\textsuperscript{travel frequency}, or at least maybe four times a year. & \small 
I travel quite often, or \hl{at least maybe four times a year}\textsuperscript{travel frequency}. & 0.00 & 0.00 \\ \hline

\end{tabular}

\caption{Examples of human and LLM highlights and the corresponding IoU and MHD scores. IoU is in the range $[0,1]$, where one indicates perfect human-LLM alignment (higher is better). As an average of embedding cosine distances, MHD is in the range $[0,2]$, where zero indicates perfect alignment (lower is better).} \label{table:human_vs_llm} 

\Description[]{A table showcasing three pairs of texts annotated by a human and the LLM, with IoU and Hausdorff distance scores for each pair.}

\end{table}

\section{Method}
Two studies were conducted in parallel with the development of LLMCode to investigate RQ 2 and RQ 3, respectively, both utilising the metrics developed in response to RQ 1. The first study observed the use of LLMCode functions by design students as part of game research projects, with the purpose of gaining an initial understanding of how designers perceive LLM-assisted qualitative coding in general, and the helpfulness of the metrics in particular. Another objective of the first study was to collect extensive empirical data on manual qualitative coding in a design context, which could be utilised as a baseline for evaluating LLM-generated codes for the same texts. The second study involved a more in-depth investigation of designers' interaction with the tool through a user interface, investigating the dynamics of influence between the participants and the tool in insight formation, and specifically what effect the developed metrics have on researcher-AI alignment.

Several types of data were collected across the two studies, including system interaction logs (both studies), survey responses (Study 1), and transcripts from think-aloud sessions with users (Study 2). The system interaction logs were quantitatively analysed to examine how insights are shaped through qualitative coding with and without AI assistance, as well as to assess the effectiveness of instructing an LLM to emulate these processes. To complement these findings, the first author conducted a reflexive thematic analysis \cite{Braun:19} of the open-ended survey responses and think-aloud transcripts, providing qualitative insights into participants' nuanced experiences of integrating LLMs into the research process.

In both studies, LLMCode was used with OpenAI’s current flagship LLM, gpt-4o \cite{OpenAI:24}, accessed via the provided API. Word embeddings were calculated with the text-embedding-3-large model from OpenAI \cite{OpenAI:24:2}.

\subsection{Study 1: Exploring LLM-Assisted Qualitative Coding in a Design Setting}\label{sec:study1}

\begin{table}[ht]
\centering
\begin{tabular}{lrlrlrrr}
\toprule
 \thead{Label} & \thead{Age group} & \thead{Gender} & \thead{Game design \\ experience (years)} & \thead{Qualitative text \\ analysis experience$^\dagger$} & \thead{$n_{coded}$} & \thead{$n_{uncoded}$} & \thead{$n_{codes}$} \\
\midrule
P1 & 30-39 & Man & 3 & TA & 48 & 162 & 17 \\
P2 & 20-24 & Prefer not to say & 1 & TA & 25 & 175 & 14 \\
P3 & 20-24 & Man & 1 & None & 97 & 102 & 8 \\
P4 & 20-24 & Woman & 1 & TA & 8 & 4 & 6 \\
P5 & 20-24 & Woman & 3 & TA & 55 & 41 & 35 \\
P6 & 25-29 & Man & 0 & None & 38 & 99 & 15 \\
P7 & 20-24 & Man & 10 & None & 32 & 76 & 15 \\
P8 & 20-24 & Man & 10 & TA & 25 & 11 & 18 \\
P9 & 20-24 & Man & 0 & None & 77 & 70 & 68 \\
P10 & 25-29 & Man & 2 & Coding & 39 & 244 & 16 \\
P11 & 20-24 & Man & 6 & TA & 23 & 228 & 6 \\
P12 & 20-24 & Woman & 2 & TA & 48 & 98 & 48 \\
P13 & 30-34 & Man & 5 & None & 14 & 43 & 10 \\
P14 & 25-29 & Man & 0 & None & 47 & 137 & 7 \\
P15 & 20-24 & Non-binary & 5 & CA & 35 & 164 & 60 \\
P16 & 20-24 & Man & 2 & Coding & 23 & 87 & 21 \\
P17 & 25-29 & Prefer not to say & 2 & None & 50 & 150 & 16 \\
P18 & 25-29 & Man & 4 & CA & 75 & 124 & 29 \\
P19 & 20-24 & Woman & 3 & TA & 42 & 39 & 17 \\
\bottomrule
\end{tabular}
\caption{Detailed information for each Study 1 participant, including statistics for the manual coding part of the study. $^\dagger$TA: Thematic Analysis; CA: Content Analysis.}
\label{tab:participants}
\Description[]{A table with participant demographics information for Study 1.}
\end{table}

The first study was conducted in conjunction with a master’s-level university course on game design research. Participation was voluntary and unincentivised. A total of 19 students took part, with three years of game design experience on average (see Table \ref{tab:participants} for detailed participant information). The study comprised two sessions with a total duration of five hours. During these sessions, participants conducted game design research projects assisted by the LLMCode toolkit, addressing research questions of their own choosing.

The first session began with an introduction to reflexive qualitative analysis and inductive coding. Participants then defined a research question and scraped relevant online discussion threads from Reddit\footnote{https://www.reddit.com}, which served as their research material. The scraping was performed using a Jupyter notebook included in LLMCode. 
Following this, participants spent the remaining two hours of the session manually coding a subset consisting of up to 200 texts from their collected corpus. A simple manual coding tool—an initial version of the LLMCode interface in Figure \ref{fig:interface} without the AI features—was provided for this task. The tool retained a log of all changes made to the material, enabling us to track the development of participants’ insights throughout the manual coding process. Table \ref{tab:participants} displays, for each participant, the number of manually annotated texts with coded segments $n_{coded}$, the number of annotated texts without coded segments $n_{uncoded}$, and the total number of distinct codes $n_{codes}$.

In the second session, participants used LLMCode functions in combination with their manually annotated set to first select a suitable example set, which was subsequently used to code the entire corpus of scraped texts. After this, participants utilised another LLMCode function to group their codes into broader themes. At the end of the second session, participants were asked to distil their key insights into a presentation slide deck, a common practice for communicating research findings in the design industry. Surveys were administered after the first session and again after the completion of the insight decks to gather data on participants’ experiences using the LLM-based tool as part of their qualitative analysis process. The surveys included free-form questions on participants' feelings towards LLM-assisted coding and theme generation, as well as their perceptions on the quality metrics and their own agency over the final insights.

\subsection{Study 2: User Evaluation of LLMCode Interface for Model Alignment}\label{sec:study2}

\begin{table}[ht]
\centering
\begin{tabular}{lrlrll}
\toprule
 \thead{Label} & \thead{Age group} & \thead{Gender} & \thead{Design experience \\ (years)}  & \thead{Types of design \\ experience$^\dagger$} & \thead{Qualitative text \\ analysis experience$^\ddagger$} \\
\midrule
P20 & 30-39 & Woman & 3 & SD & TA \\
P21 & 25-29 & Woman & 2 & SD & Coding \\
P22 & 20-24 & Man & 7 & GD & CA \\
P23 & 25-29 & Woman & 8 & UX, ID & TA \\
P24 & 25-29 & Prefer not to say & 4 & PD, SD & CA \\
P25 & 25-29 & Woman & 8 & UX, UI & UR \\
P26 & 25-29 & Man & 6 & GD, GrD & TA \\
\bottomrule
\end{tabular}
\caption{Detailed information for each Study 2 participant. $^\dagger$SD: Service Design; GD: Game Design; ID: Industrial Design; PD: Product Design; GrD: Graphic Design. $^\ddagger$TA: Thematic Analysis; CA: Content Analysis; UR: User Research.}
\label{tab:participants2}
\Description[]{A table with participant demographics information for Study 2.}
\end{table}

The second study involved sessions where participants with design backgrounds were asked to use the LLMCode interface to first manually code some research material, and then iterate on their chosen examples to improve the model's performance as measured by the two metrics. Seven participants were recruited for the study through a combination of direct and indirect outreach within the authors' professional and academic networks. This approach was chosen to ensure access to individuals from a wide range of design expertise and perspectives (see Table \ref{tab:participants2} for details). To enhance the contextual validity of the study, participants were encouraged to bring their own data for analysis, provided this was permitted by relevant guidelines or regulations. For the four participants unable to do so, transcription data from three interviews conducted by service designers was provided along with a fictional service design task.

The participants followed a think-aloud protocol while completing the tasks. The sessions were conducted remotely via video call, with participants sharing their screens throughout the study. Each session lasted approximately 90 minutes,  divided evenly between two tasks: manually coding data and iteratively refining the example sets provided to the model. At the conclusion of the first phase, any unannotated data was discarded, and participants were asked to select an initial example set to provide to the model that is representative of their coding. The remaining texts were then coded by the model, with its annotations displayed alongside the participants’ annotations for comparison.

During the second phase, participants were instructed to complete two additional iterations of example selection and LLM-assisted coding using the tool's features, with the objective of improving the code quality metrics. In addition to modifying the example sets, participants were encouraged to refine their own annotations and include additional prompt instructions for the model. Due to the tool's complexity, participants were guided through each feature as it was introduced and encouraged to ask questions as needed. This approach ensured participants could engage effectively with the tool and focus on the analysis process.

\section{Results}

\subsection{(Superficial) Gains in LLM Performance Through In-Context Learning}
\label{sec:results-in-context-learning}

\subsubsection{In-Context Learning Performance}

Most state-of-the-art qualitative coding systems based on LLMs rely on in-context learning, which assumes that the model can learn coding patterns from a sufficient number of human-coded examples. To investigate this assumption, we examine how the performance of the model—measured by the IoU and MHD metrics—improves as the number of few-shot examples increases.

For this analysis, we selected data from 11 participants in Study 1., who had annotated sufficiently large manually coded datasets, with $n_{coded} \geq 38$, $n_{uncoded} \geq 38$, $n_{codes} \geq 6$. For each participant, we compared their manually annotated texts with the annotations generated by the LLM. For this purpose, the annotated data for each participant was divided into two groups: one set of examples was used to teach the model, and the other set was used to evaluate its performance. The examples used to teach the model included both positive examples—texts containing at least one coded section—and negative examples—texts with no annotations. To account for variations in coding styles and corpora, we selected equal numbers of positive and negative examples for training.

To understand how the model's performance changes with varying amounts of input data, we evaluated its performance at different example set sizes. The examples were selected from each participant's coding logs in chronological order, reflecting how their insights evolved over time. By doing this, we accounted for the fact that participants might refine or change their annotations as they progressed.

\begin{figure}
    \centering
    \includegraphics[width=0.5\linewidth]{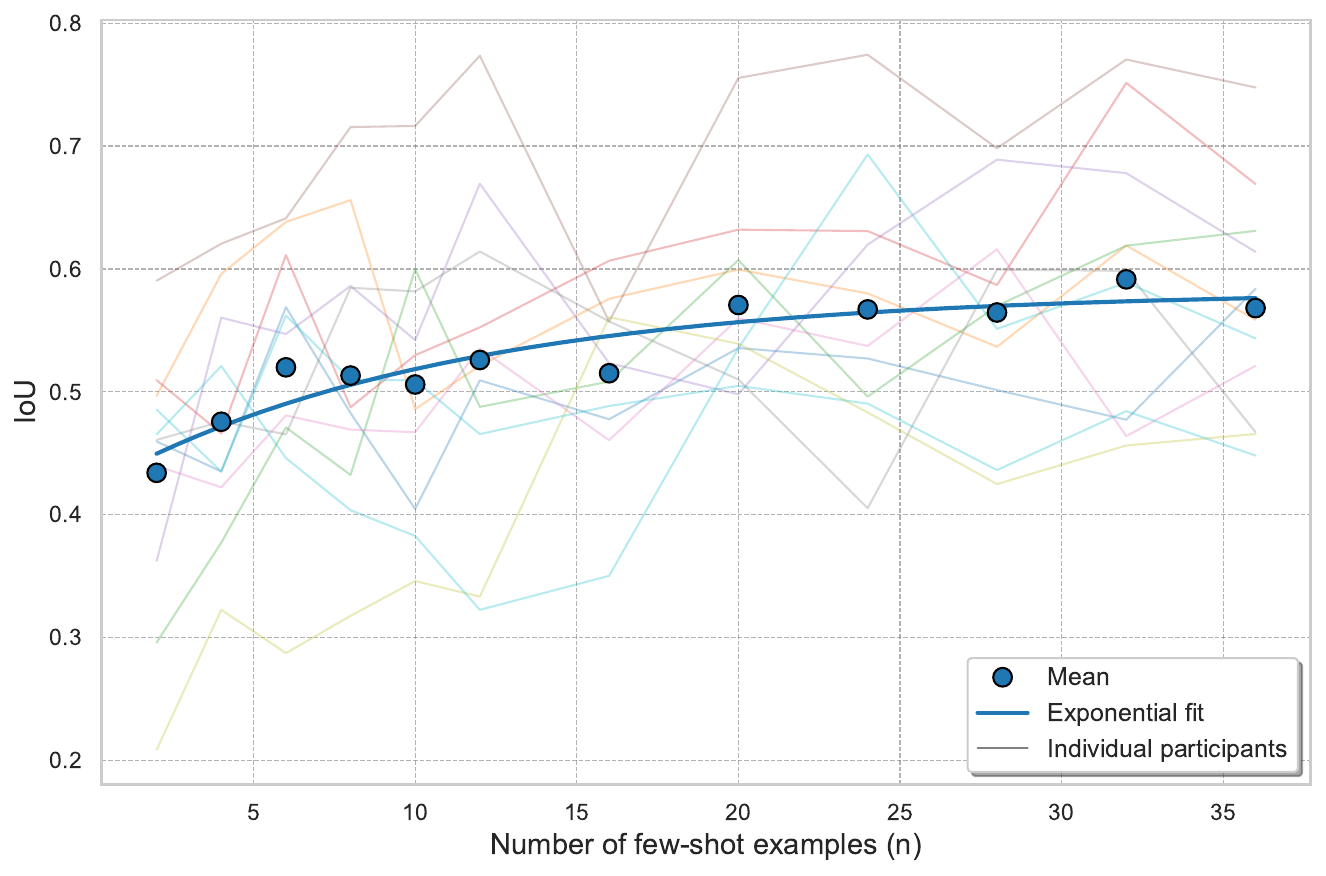}
    \caption{Mean IoU plotted against the number of few-shot examples included in the coding prompt (Study 1, N=11). The metric increases as the number of few-shot examples increases, indicating improved alignment between the LLM and human annotations.}
    \label{fig:iou}
    \Description[]{The figure shows a scatterplot of points and an increasing exponential curve fitted to them. The figure describes IoU development with the number of examples. Behind the main curve there are several faint thin lines, showing how the metric changes for individual participants.}
\end{figure}

Figure \ref{fig:iou} illustrates how IoU, averaged across all texts, changes with the number of examples for individual participants and as an average across all included participants. The smaller lines, representing individual participants, reveal significant variability in the model’s performance depending on the selected examples. Since examples in this analysis were selected automatically, erroneous or non-representative examples may significantly lower the model’s performance. In practical applications, we recommend that researchers manually select the most representative examples using a human-annotated validation set, in a method that is investigated in Study 2.

We observe an increasing trend in the mean IoU values averaged across all participants. This suggests that through in-context learning, the model can to some degree emulate an individual researcher’s perspective on what content is interesting or important. To model this behaviour, an asymptotic exponential growth curve
is fitted to the data using the least squares method. The choice of this function is motivated by its previous use in learning curve analysis \cite{Vianna:24} as well as the expected asymptotic learning behaviour of the model as the size of the prompt approaches the GPT-4o model's context window of 128,000 tokens \cite{OpenAI:24}. The maximum length for an individual prompt in this study was 391,414 characters or approximately 98,000 tokens \cite{OpenAI:24:3}.

\begin{figure}
    \centering
    \includegraphics[width=0.5\linewidth]{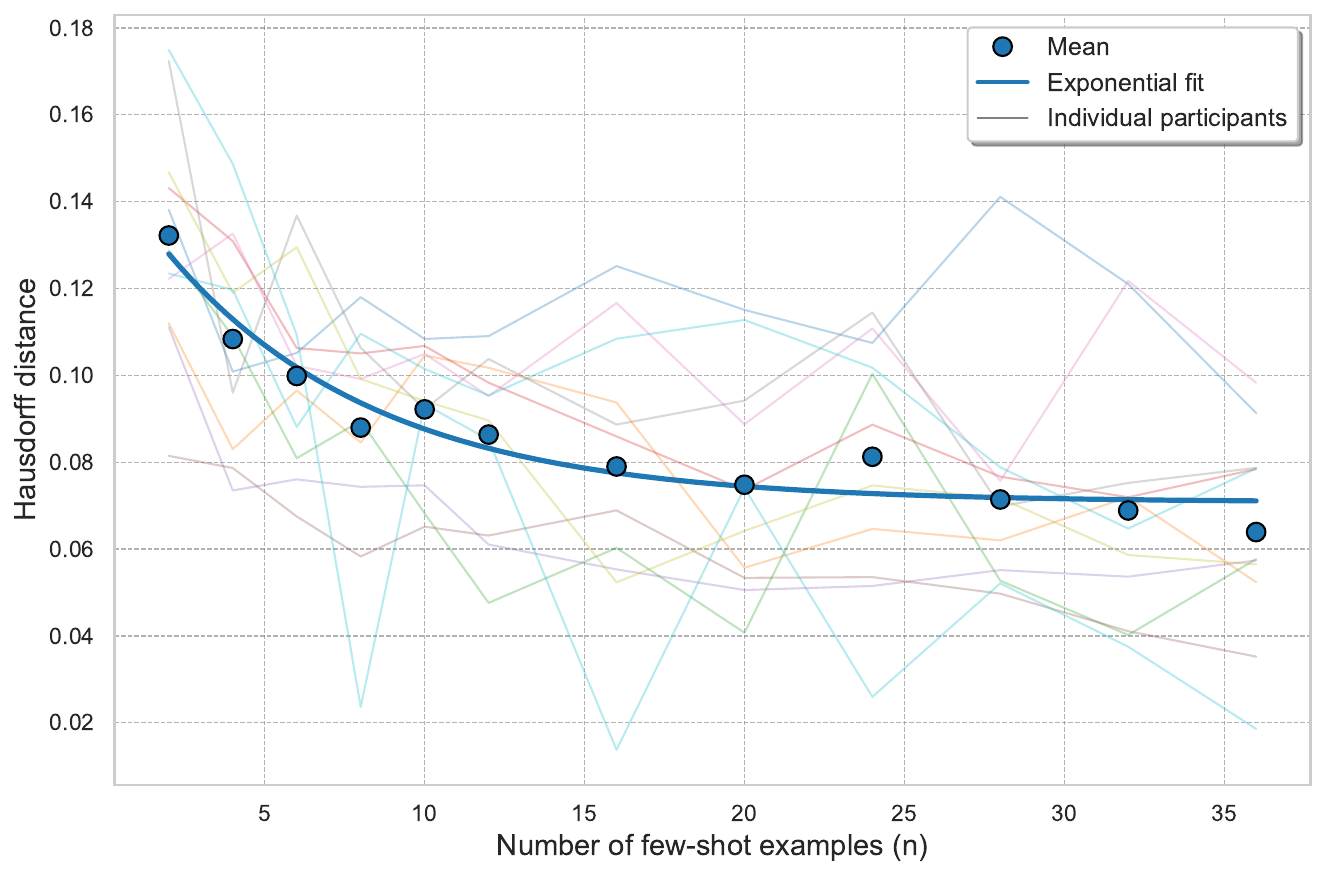}
    \caption{Mean MHD plotted against the number of few-shot examples included in the coding prompt (Study 1, N=11). The metric decreases as the number of few-shot examples increases, indicating improved alignment between the LLM- and human-annotated codes.}
    \label{fig:hausdorff}
    \Description[]{The figure shows a scatterplot of points and a decreasing exponential curve fitted to them. The figure describes MHD development with the number of examples. Behind the main curve there are several faint thin lines, showing how the metric changes for individual participants.}
\end{figure}

\begin{figure}
    \centering
    \includegraphics[width=0.35\linewidth]{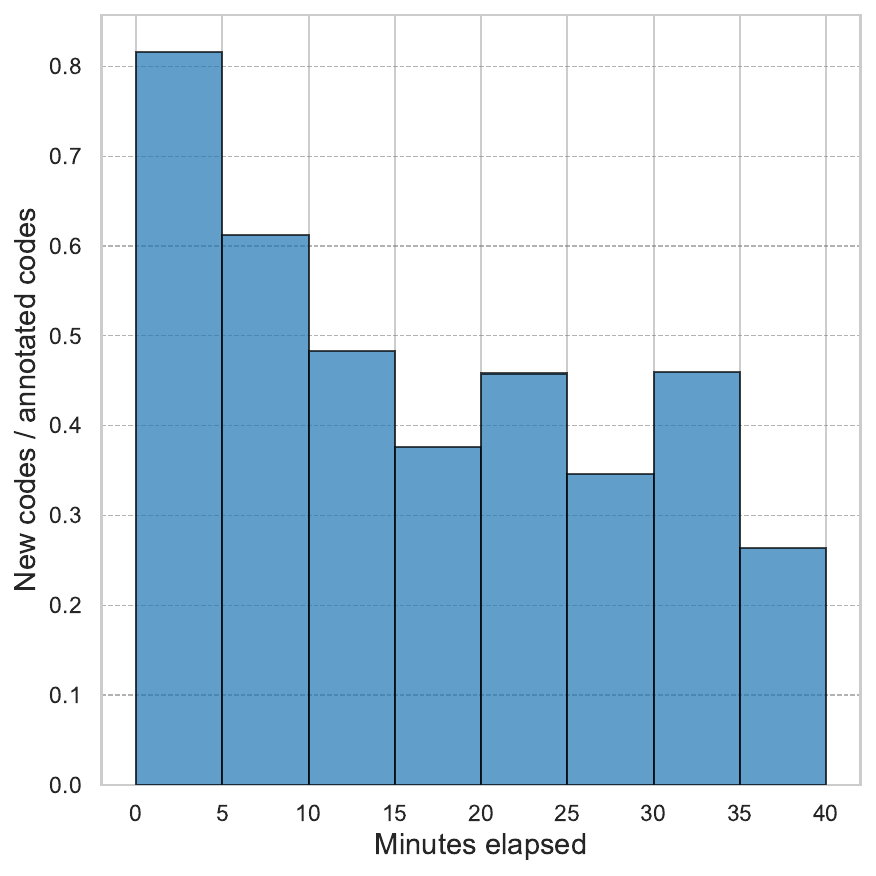}
    \caption{New codes as a fraction of all annotated codes in the given time frame (Study 1, N=15). The rate at which new codes are discovered decreases over time as the codebook becomes saturated.}
    \label{fig:newcodes}
    \Description[]{The figure shows a decreasing barplot showing how the fraction of new codes decreases with time annotated.}
\end{figure}

A similar trend is observed for the MHD metric in Figure \ref{fig:hausdorff}, where the mean score decreases as more examples are provided, indicating improved alignment between the LLM- and human-annotated codes. However, certain codes may already be present within the selected few-shot examples. In these cases, MHD primarily reflects the model's deductive coding capabilities—specifically, its ability to apply existing codes to relevant texts.

As shown in Figure \ref{fig:newcodes}, the rate at which participants discover new codes declines over time, evidenced by the decreasing fraction of newly identified codes as more time is spent analysing the data. This suggests that the improvement in the MHD metric may not necessarily indicate a deeper emulation of the researcher's analytical perspective. Instead, it could result from the examples becoming more representative of the dataset over time, as fewer new codes are introduced. Consequently, the metric may be capturing the convergence of examples and inference data rather than the model’s ability to fully reflect the researcher’s interpretive lens.

\subsubsection{Finding New Insights Beyond the Examples}

\begin{figure}
    \centering
    \includegraphics[width=0.5\linewidth]{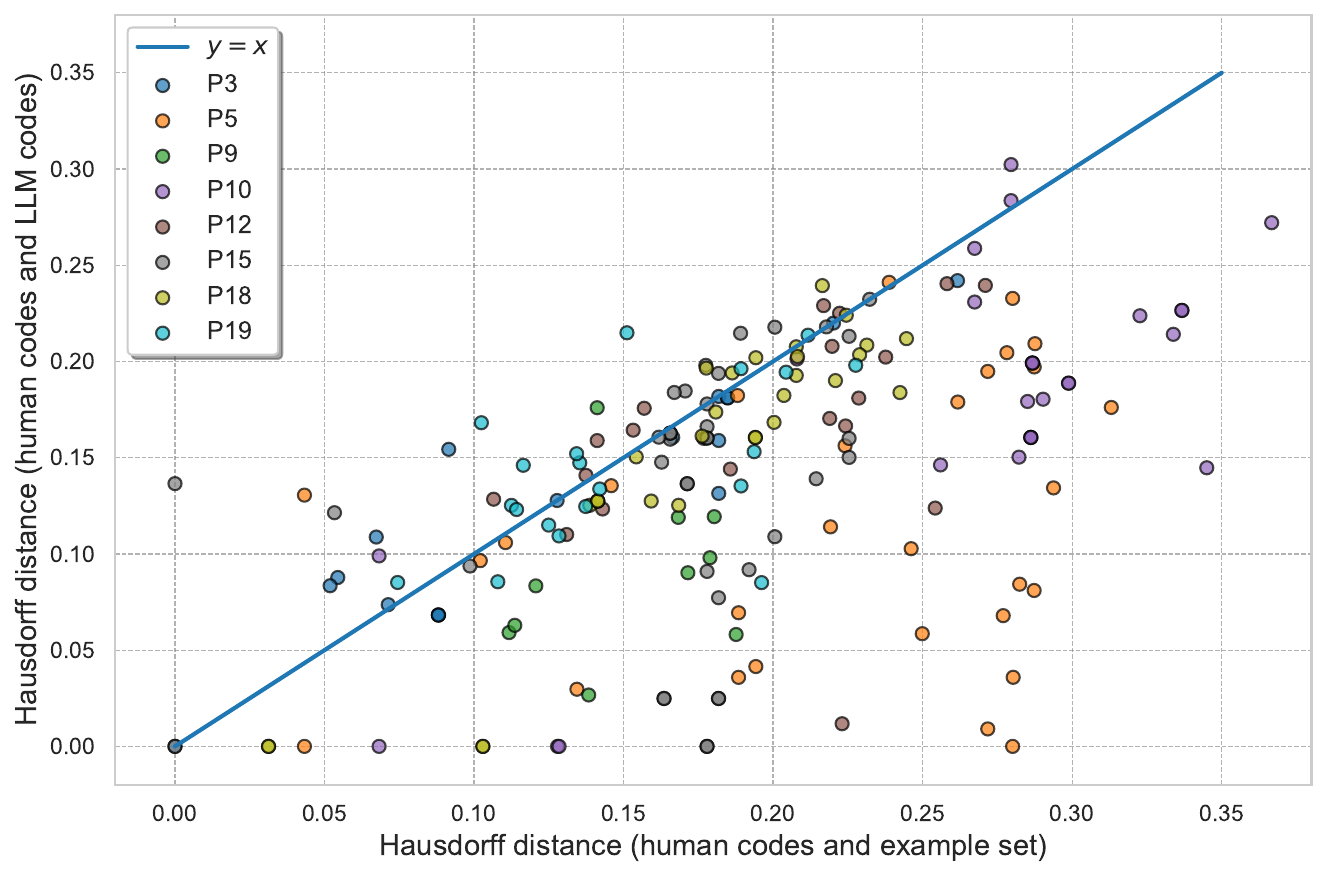}
    \caption{Scatter plot showing the correlation (Pearson coefficient = 0.53) between the similarity of human-annotated codes to the example set and the model’s performance (Study 1, N=8). Higher dissimilarity generally leads to lower alignment, though some cases show the model successfully extrapolating beyond the examples. A line of equality $y = x$ is added to illustrate how the model rarely performs worse than what is expected based on its examples' similarity to the texts' underlying codes.}
    \label{fig:correlation}
    \Description[]{The figure shows a scatter plot of points displaying moderate positive correlation. Each point is coloured by a participant. A line y = x is plotted in the middle of the figure.}
\end{figure}



To ascertain whether the model can extrapolate the researcher’s lens beyond the provided examples, we 
conduct an additional analysis using data from Study 1. For each participant, their final codes are clustered into five distinct groups via K-Means clustering on the code embeddings. This method aims to capture broader themes within the data by grouping codes that are semantically similar while maximising differences between groups, as determined by the semantic information encoded in the embeddings.

To evaluate whether the model can uncover new insights beyond a familiar theme, it is provided with texts containing codes from only one of these clusters as examples. The model is then tested on texts annotated by the human researcher with a more diverse range of codes spanning different clusters. For each processed text, we examine the correlation between the similarity of the text's human-annotated codes to the example set (measured by the MHD between the human codes and the example set) and the model's performance on that text (measured by the MHD between the human codes and the LLM-generated codes).

The results, presented visually in Figure \ref{fig:correlation}, reveal a Pearson correlation coefficient of 0.53, indicating a moderate positive correlation. This suggests that as the dissimilarity of the human annotations to the example set increases, the model’s performance generally declines, meaning that its output diverges from the human output. Interestingly, some cases in the lower right-hand corner of the plot show instances where the model successfully generates codes similar to those of the human researcher despite having been given examples that differ significantly. This indicates that the model may, in certain instances, be capable of extrapolating beyond the provided examples to identify patterns aligned with the researcher’s lens.

However, the overall trend suggests that the model performs better on texts that are more similar to its example set. This finding underscores the model’s limitations in generalising across broader perspectives and inductively generating new ones, highlighting the importance of providing diverse and representative examples to enable its alignment with the human researcher. This finding, together with the declining rate of new code discovery over time, supports a practical workflow where researchers begin by manually coding a subset of texts to generate an initial codebook, which the automated system can subsequently apply to larger datasets.

\subsection{Human Oversight Can Lead to Improved Alignment}
\label{sec:study-2-quant}

\begin{figure}[ht]
    \centering
    \begin{subfigure}{0.48\linewidth}
        \centering
        \includegraphics[width=\linewidth]{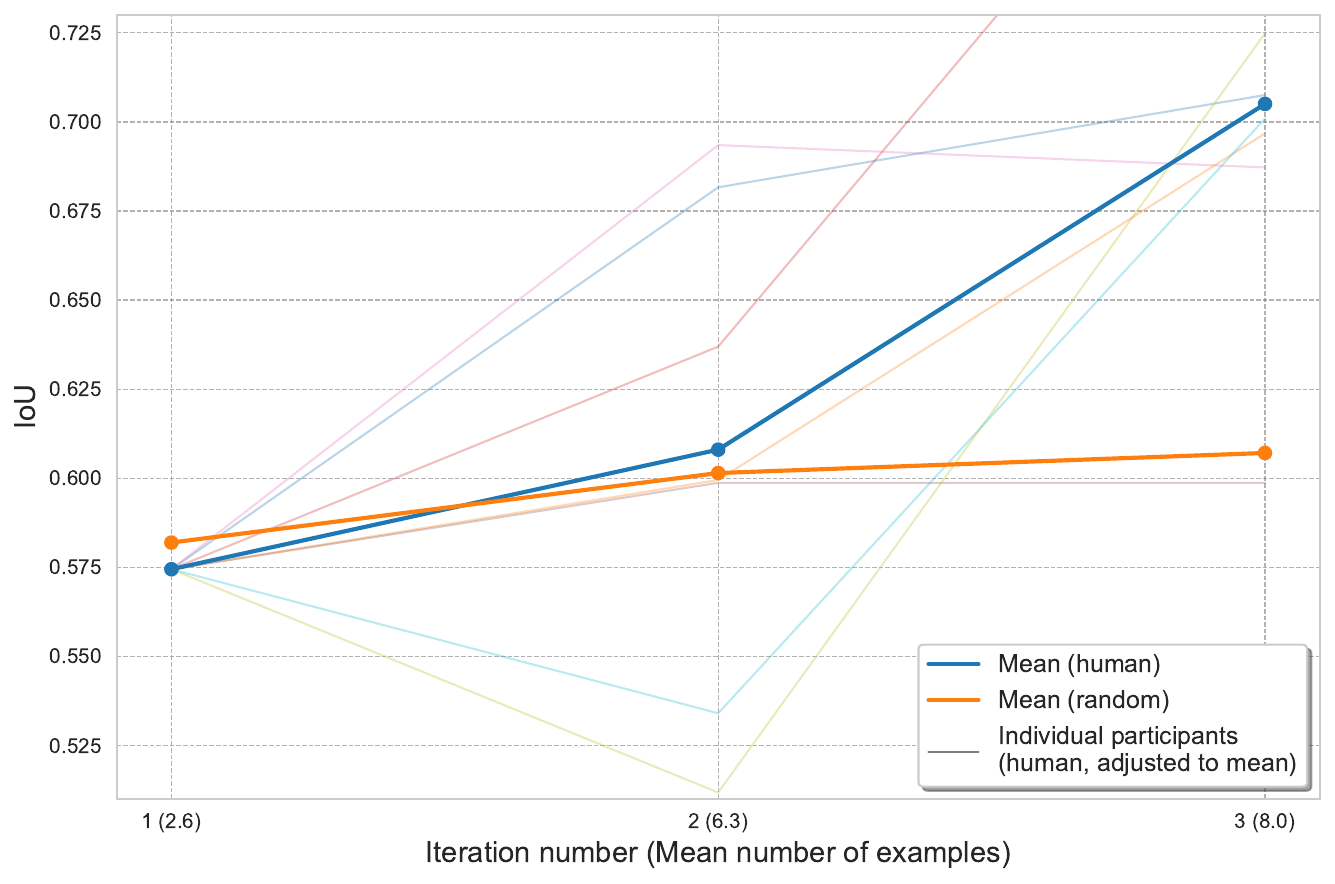}
        \caption{IoU}
        \label{fig:iou2}
    \end{subfigure}
    \hfill
    \begin{subfigure}{0.48\linewidth}
        \centering
        \includegraphics[width=\linewidth]{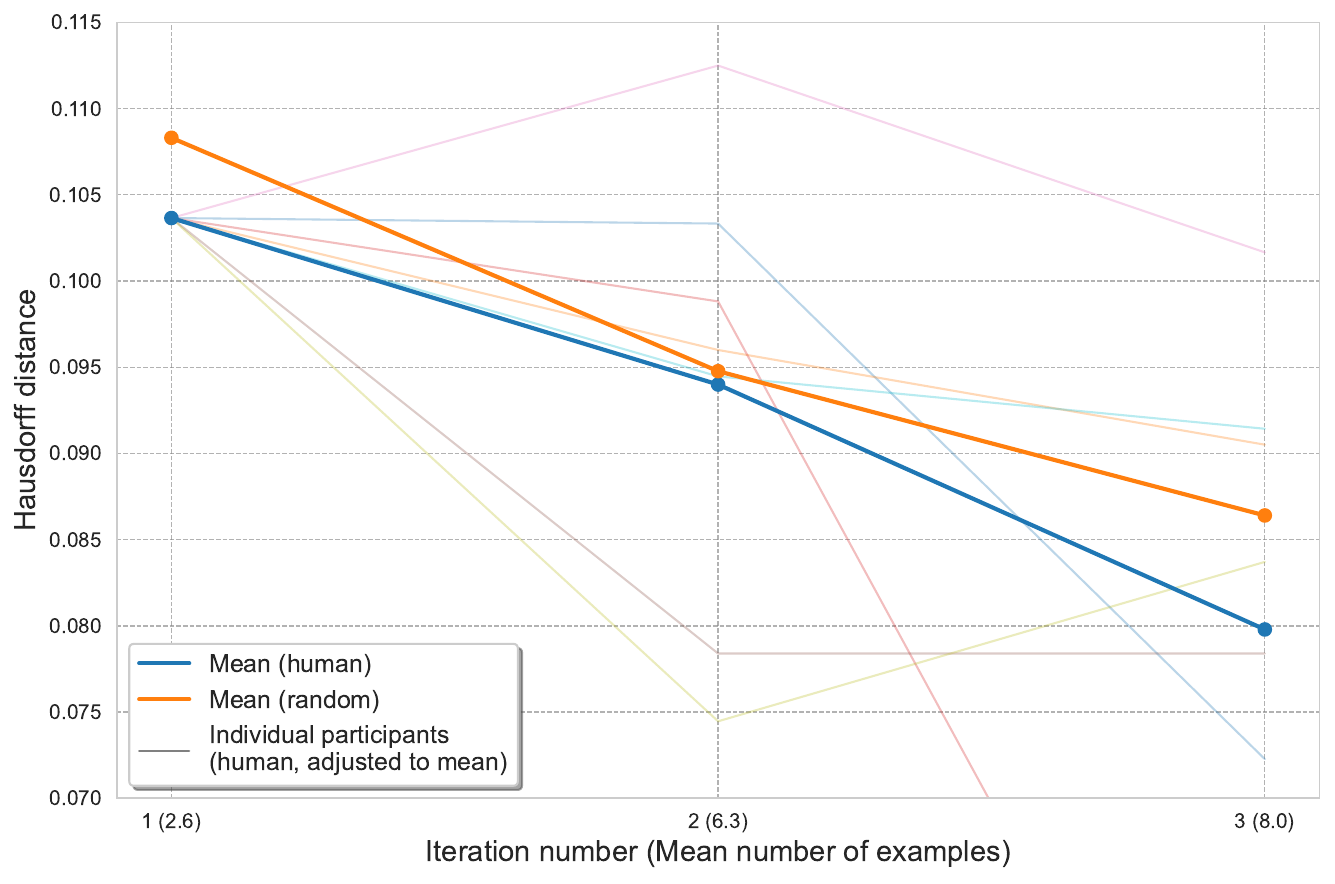}
        \caption{Modified Hausdorff distance}
        \label{fig:hausdorff2}
    \end{subfigure}
    \caption{The development of (a) IoU and (b) Modified Hausdorff Distance as participants in Study 2 manually iterated on their few-shot example sets (human), compared to a random sampling baseline with equal numbers of positive and negative examples (random). Curves for individual participants are adjusted to match the initial mean, for visualisation purposes.}
    \label{fig:manualselection}
    \Description[]{The figure shows two subfigures, each with two similar line plots in blue and orange. The blue line shows slightly better metric improvement in both, corresponding to the system's performance with human assistance.}
\end{figure}

To evaluate whether researchers can improve the model's alignment by iteratively refining their few-shot example sets using our tool, we compared their results across the iterations in Study 2 to a random selection baseline, as shown in Figure \ref{fig:manualselection}. Since the model’s performance generally improves with the number of provided examples, the random baseline was designed to match the number of examples selected by participants at each iteration, maintaining an equal balance of positive (coded) and negative (uncoded) examples. To account for variability in random selection, the baseline results were averaged over five randomly generated example sets for each participant and iteration.

The findings suggest that iterating on the examples using our tool can lead to better results, particularly for the IoU metric. In the first iteration, participants lacked insight into the model’s behaviour from prior iterations, requiring them to make initial guesses about which examples would be most representative. This likely explains why their scores in the first iteration were similar to, or slightly lower than, the random baseline. In subsequent iterations, participants gained access to the model’s annotations from the previous iteration, enabling more informed adjustments. This iterative process resulted in more significant improvements in IoU scores compared to the random baseline, particularly by the third iteration.

However, the data from this study is limited, and similar trends were less conclusive for the MHD metric. Despite this, the increasing rate of improvement in the third iteration aligns with qualitative observations from the user study, where participants initially struggled to grasp the rationale behind example selection but eventually found the tool valuable in refining their alignment.

\subsection{Thematic Analysis}

The thematic analysis is based on the responses to open-ended survey questions from Study 1 and 428 minutes of transcribed think-aloud recording data from Study 2, resulting in 149 quotations and 63 codes that were iteratively refined and grouped under two overarching themes. In line with our study topic, we adopted a reflexive approach to thematic analysis, allowing themes to to be determined organically through iterative coding rather than adhering to a structured theoretical framework \cite{Braun:19}. This approach enabled us to capture the richness and complexity of participants’ experiences, providing qualitative insights into how they interacted with the tool and sought to align their understanding with its analysis.

\subsubsection{If AI Tells Me to Change It...}
\label{sec:ai-impact}

Participants found that ranking texts by the highlight and code similarity metrics allowed them to efficiently identify cases where the model’s coding diverged from their own annotations, particularly in the context of the full LLMCode user interface. Where in the context of the initial study's Notebook environment some participants had perceived the metrics as "unintuitive" (P4) and "uncontrollable" (P10), the ranking feature in the user interface provided a straightforward way to focus on areas of disagreement, making it easier to evaluate the model’s performance, find better examples for the model, and keep track of the model's development.

A key complaint on the system's coding among participants was that its annotations were often uninformative or lacked contextual understanding, rendering them meaningless in relation to the research question. One participant noted that the model's inability to distinguish "meme comments” (P10) made it difficult to utilise in the context of online forum data, which often includes niche humour that may entirely shift the meaning of a text. Another participant observed that the AI “used very vague codes that sometimes overlapped a lot.” (P2) Similarly, another participant expressed frustration with the AI assigning codes to what they perceived as trivial responses: “When they’re just very short answers like ‘yes,’ [...] they’re not that interesting.” (P26) These comments highlight the model’s inability to consistently produce nuanced or contextually appropriate annotations.

Despite these limitations, participants often found that the system’s annotations prompted them to reconsider and refine their own coding. Participant P25, for example, initially coded a segment as “background info” but removed it after noticing that the model had not annotated the same segment. In a subsequent iteration, the model assigned the code “Participant Background” to the same segment, prompting the participant to adopt this suggestion back into their own coding. Reflecting on this, they admitted being influenced by the AI’s annotations, revealing how the system could subtly shape research insights.

Furthermore, the alternative perspective provided by the AI sometimes led participants to question their own interpretations. While this divergence could feel disorienting, it occasionally appeared to encourage deeper introspection and re-framing of their analytical lens. The model’s tendency to introduce slightly different wordings, for instance, occasionally sparked confusion but also led to new insights. One participant (P23), while comparing AI-generated codes related to a game, reflected: "There are a few codes that are new, 'surprise' and 'story'. But I'm thinking, how useful are they? They are very close to novelty and lore, I would say. But, I guess lore is more about, like, facts of the world and the stories about, like, how things happen." These moments show how the system's suggestions could prompt participants to reconsider and refine their understanding of nuanced concepts.

Participants’ attitudes toward the AI’s output varied. Some preferred to stick to their original insights, as one explained: “I’m not sure what to do with the AI’s annotations—should I incorporate them into my coding framework or not? I assume this tool is meant to assist me in my work, but since I’ve already coded this myself, I’m not sure if I want to include them.” (P20) These reflections illustrate the tension between embracing new perspectives and over-attributing meaning to the AI’s output, which participants sometimes found inconsistent. As one noted, trusting the AI’s annotations could feel like “a leap of faith” (P26), highlighting the complex interplay between human judgment and AI influence in the tool's collaborative analysis.


\subsubsection{Let Me Explain My Examples}

The primary interaction method for teaching the model—designed around the in-context learning paradigm and involving iterative selection of representative few-shot examples—proved challenging for many participants, especially those with limited experience using LLMs beyond standard chat interfaces. Participants often expressed confusion about how to select and use examples to train the model, with several describing the “make example” feature as vague or difficult to interpret. As one participant put it: “This is a bit unclear to me […] I’m making this an example, but an example of what exactly?” (P20) 

Many participants appeared to intuitively associate examples with the model’s perceived \textit{actions} in relation to the human-coded texts. One participant explained: “It was able to use the codes I gave it and subdivide them into individual smaller parts: I think that’s what makes this example particularly good. So if I put this comment as an example, maybe I would see more of this subdivision \textit{behaviour}.” (P23) This suggests that participants viewed examples as a means of reinforcing desirable changes between the human- and LLM-generated annotations, rather than as independent demonstrations of appropriate outputs. This framing of examples as actions rather than outputs shaped how participants interacted with the system and contributed to their difficulty grasping the intended purpose of example selection.

Another, possibly related, challenge stemmed from participants’ inability to specify why an example was chosen. Some expressed frustration with the lack of options to communicate the reasoning behind their choices. For instance, P26 deliberated over whether to include a poorly annotated text in the example set, saying: “I wonder if this will just make [the model] confused. If I now put this as an example, will it [understand what I mean]? I wish there was an option here to mark that ‘You’ve misunderstood the text.’” To be able to interpret such feedback, the model would necessarily have to learn from a more multidimensional history of both correct and mistaken annotations, rather than relying solely on a set of static, exemplary annotations. This reflects participants’ desire for the system to engage with the reasoning behind their selections rather than treating examples as isolated data points. Similarly, P20 wished for the ability to “qualitatively describe” why an example was selected, while P23 longed for text-level prompting to build upon their own annotations: “Here's an emotion: now, go figure it out, machine!”

On the other hand, many participants faced difficulties in articulating the instructions they wished to communicate to the model. For example, P22 struggled to craft a prompt that would encourage the model to generate more nuanced emotional codes: “Similar to the AI, I have no idea about human emotion, so I don’t know what to write although I want the correct answer […] because the definitions are a little bit ambiguous.” This comment reflects the interpretive nature of RfD, where correct answers often coexist with ambiguous definitions, leaving researchers to rely on what P26 described as their “gut feeling” to determine the validity of insights. In such situations, the use of illustrative examples may provide the simplest way forward, as P22 ultimately chose to do.



\section{Discussion}

Based on our results, finding a path towards true researcher-LLM alignment remains a challenge. In this paper, we focused on methods based on in-context learning, given that this is the standard for LLM tools in this domain. The following sections discuss our findings, first relating to our empirical study of the model's capabilities for in-context learning in this task, and after that, to our study of human-AI interaction in facilitating the learning process. We conclude by discussing best practices for those looking to use LLMCode in practical qualitative research settings.

\subsection{LLMs May Only Emulate Surface-level Analysis}
\label{sec:llm-analysis-discussion}

Our empirical findings detailed in Section \ref{sec:results-in-context-learning} indicate that increasing the number of few-shot examples can enhance alignment between an LLM and a human researcher’s analytical perspective, as observed in the upward trend of the IoU metric and the decreasing Modified Hausdorff distance. However, closer examination suggests that these gains may primarily reflect the model’s ability to adapt or reuse codes present in the training examples, rather than provide evidence of deeper, inductive reasoning. The plateauing rate at which participants discovered new codes further underscores that the model’s improved performance may be largely attributable to converging on a codebook that is already stabilising, rather than genuinely learning novel or more nuanced insights.

The model’s performance was correlated to a significant degree with the representativeness of the examples it was given. Although there were instances where the model succeeds in coding texts whose themes differ significantly from its initial examples, these cases were the exception. Overall, such observations reinforce that the model’s apparent learning remains confined to the surface features supplied by the human-annotated examples. The challenge of true interpretive alignment—where an LLM would spontaneously identify, refine, or propose new codes that mirror a researcher’s emergent understanding—remains largely unmet.

Prior work by \citet{Ashwin:23} has proposed that LLMs be used for extending human analysis to larger corpora, and our study clarifies when this application is likely to be effective. In contexts where the codebook is relatively established and the new data remains thematically close to previously annotated examples, the model’s performance can scale to facilitate broader analysis. However, if truly novel themes or codes arise in the data, the model may struggle to adapt in a meaningful and contextual manner, highlighting the need for ongoing human oversight and potential recoding when stepping beyond the scope of the original examples.

\subsection{Designing Tools for Researcher-AI Alignment in Qualitative Analysis} 

Our findings reveal both the promise and complexity of integrating LLMs into interpretive research processes. Participants appreciated how our tool could quickly highlight areas of disagreement, but many struggled to improve this alignment through the proposed method of example selection. 

\subsubsection{Disconnect Between Chat Mental Models and In-Context Learning}

A core challenge in our user studies was the disconnect between the mental models participants had formed through typical chat interactions with LLMs and the fundamentally different requirements of few-shot in-context learning. In conventional chat-like interactions, users iteratively refine the model’s behaviour by responding directly to its actions—correcting mistakes, prompting for clarifications, and building on previous turns. Users in our study carried this assumption into their interactions with LLMCode, naturally seeing corrections and prompts as feedback that would accumulate over time.

However, in-context learning as implemented in our tool worked differently. Rather than creating a long-term conversational memory, users had to manually select and compile a static set of annotated examples, all of which were submitted to the LLM at once to shape its coding behavior. The system did not automatically “remember” prior turns or corrections; it also provided no default interface for explaining \textit{why} an example was correct or incorrect. This design led many participants to focus on “rewarding” perceived improvements by selecting examples where the AI seemed to have done something new or valuable—an interesting finding that underscores how interface design can strongly shape user expectations and behaviour.

Notably, the participant who adapted most fluidly to this static example-based paradigm was someone who had prior experience building LLM applications, suggesting that familiarity with in-context learning significantly reduces confusion. Future iterations of few-shot learning interfaces might therefore need more explicit guidance or visual cues to reinforce how in-context learning differs from conversational AI. Alternatively, such tools might allow users to provide more flexible examples of how the coding should \textit{change}—for instance, showing the AI’s initial annotation and then demonstrating a “corrected” version—which would align better with users' intuitive teaching practices.

\subsubsection{The Reciprocal Influence of AI Suggestions on Human Analysis}

Although participants faced obstacles when trying to teach the model via few-shot examples, our findings also illuminate a reverse flow of influence. Many participants willingly incorporated new perspectives introduced by the LLM. For instance, when participants noticed that the AI generated an unfamiliar code, they sometimes adopted that code as an additional lens for interpreting the data. In these instances, the model’s output prompted users to revisit or refine their original coding frameworks. This openness often stemmed from curiosity—participants wanted to see if the AI might surface patterns they had overlooked. At the same time, they exercised caution, especially in cases where the model’s recommendations clashed with their domain knowledge or research goals.

Ultimately, these findings suggest that a well-designed tool could not only help align AI outputs with a user’s perspective, but also empower users to see their data from alternative angles, blending both human and machine insights into a richer analytical process. However, in light of our previous discussion in Section \ref{sec:llm-analysis-discussion}, trusting LLMs as research partners requires further research into how reliably they can approximate the interpretive reasoning of human researchers. If it remains beyond current capabilities to emulate real-world researchers, then it becomes even more critical to understand the model’s inherent perspectives and biases, and how these might shape or constrain the insights it produces. In addition, there is a need to investigate appropriate reliance on such tools more deeply, following the approach suggested by \citet{jacovi:21}, which examines the relationship between model correctness and user reliance in interactive scenarios. Through such investigations, we can better establish the conditions under which LLM-driven tools are beneficial to qualitative research, ensuring that their adoption does not compromise the reflexive and context-rich nature of design inquiry.

\subsection{Training, Validation, and Test Data}
\label{sec:sufficient}


The example iteration approach introduced in Section \ref{sec:interface} and evaluated in Study 2 is susceptible to \textit{overfitting}, which is an important consideration for using LLMCode in high-stakes qualitative studies, such as those conducted in academia where rigorous quality standards are expected. Overfitting, a common issue in machine learning \cite{dietterich1995overfitting}, occurs when a model performs exceptionally well on the data it was fine-tuned with but fails to generalise to unseen data. In the context of LLMCode, this would mean high alignment scores on the annotated texts used during the example iteration process, but potentially poorer performance on texts excluded from this set.

To account for overfitting, standard machine learning practices should be followed, dividing data into \textit{training}, \textit{validation}, and \textit{test} datasets. The training set is typically used to fit a model’s parameters—which is not applicable here since LLMs are pre-trained—while the validation set serves to optimise hyperparameters, in this case the prompt and examples. The test set, crucially, evaluates how well the model and chosen hyperparameters generalise to new data outside the validation process.

In Study 2, we did not employ a separate test set due to the time-intensive nature of manual coding and to avoid participant fatigue. As a result, while the score improvements reported in Section \ref{sec:study-2-quant} signal that the tool helps researchers align their analysis within the validation set, these scores may not fully reflect the true human-AI alignment. A further complication in measuring the true alignment is our finding that the iterative process of refining prompts and examples can influence the researcher’s coding approach itself (see Section \ref{sec:ai-impact}), thereby shifting the baseline against which alignment is assessed.

For those employing LLMCode in rigorous academic qualitative research or high-stakes RfD applications, we recommend the following steps to verify human-LLM alignment:
\begin{enumerate}
    \item \textbf{Manually code a separate test set.} The test set should be coded only after completing the iterative refinement of prompts and examples based on the validation set. This avoids introducing bias from changes in the researcher’s coding approach during the iteration process.
    \item \textbf{Report alignment metrics.} Calculate and report IoU and Modified Hausdorff distance metrics for both the validation and test sets, ensuring transparency about the model’s generalisation capabilities.
    \item \textbf{Manually inspect worst-case performance.} Include a table of worst-case results by sorting the test set based on the alignment metrics.
\end{enumerate}
By following these procedures, stakeholders can evaluate human-LLM alignment both qualitatively and quantitatively, ensuring the robustness of insights derived from the LLM-assisted analysis.



\section{Conclusion}

Based on the qualitative feedback from our studies, the proposed IoU and MHD metrics appear to be suitable indicators for the quality of LLM-assisted coding, as measured by its output's similarity to a researcher's annotations. In this study, we applied the metrics in two distinct ways. Firstly, we used the metrics to investigate the capability of a state-of-the-art LLM in emulating an individual designer's perspective on data through in-context learning. Secondly, the metrics were integrated into the interactive LLMCode coding tool, which was employed in a user study investigating the ways and extent to which AI assistance shapes the insights generated through RfD.

These studies generated important insights two sides of the appropriate reliance equation: firstly, the technology's trustworthiness, and secondly, users' willingness to rely upon it. Our results indicate that while the model could learn to capture similar surface patterns, it does not always appear to be capable of emulating the researcher's deeper interpretive lens over the data from examples alone. On the other hand, researchers interacting with the tool were willing to both educate the model on their own perspective as well as to adapt their analysis the model's output, sometimes to a significant degree.

These findings underscore the complex relationship between AI capabilities and human judgment in qualitative analysis, revealing how AI can both complement and challenge human interpretations, while emphasising the importance of aligning technological assistance with the nuanced demands of reflexive analysis. While the metrics provide a useful means to evaluate alignment, they also underscore the need for careful consideration of how LLM-based tools are integrated into RfD workflows. Future work should focus on further investigating and enhancing the interpretive capabilities of LLMs while exploring novel interfaces that encourage fluid collaboration between researchers and these models.

\bibliographystyle{ACM-Reference-Format}
\bibliography{references}


\begin{thebibliography}{47}


\ifx \showCODEN    \undefined \def \showCODEN     #1{\unskip}     \fi
\ifx \showDOI      \undefined \def \showDOI       #1{#1}\fi
\ifx \showISBNx    \undefined \def \showISBNx     #1{\unskip}     \fi
\ifx \showISBNxiii \undefined \def \showISBNxiii  #1{\unskip}     \fi
\ifx \showISSN     \undefined \def \showISSN      #1{\unskip}     \fi
\ifx \showLCCN     \undefined \def \showLCCN      #1{\unskip}     \fi
\ifx \shownote     \undefined \def \shownote      #1{#1}          \fi
\ifx \showarticletitle \undefined \def \showarticletitle #1{#1}   \fi
\ifx \showURL      \undefined \def \showURL       {\relax}        \fi
\providecommand\bibfield[2]{#2}
\providecommand\bibinfo[2]{#2}
\providecommand\natexlab[1]{#1}
\providecommand\showeprint[2][]{arXiv:#2}

\bibitem[AI(2022)]%
        {OpenAI:22}
\bibfield{author}{\bibinfo{person}{Open AI}.} \bibinfo{year}{2022}\natexlab{}.
\newblock \showarticletitle{Introducing ChatGPT}.
\newblock \bibinfo{howpublished}{https://openai.com/index/chatgpt/}.
\newblock  (\bibinfo{date}{30 Nov} \bibinfo{year}{2022}).
\newblock
\newblock
\shownote{Accessed: 2025-01-08}.


\bibitem[Ashwin et~al\mbox{.}(2023)]%
        {Ashwin:23}
\bibfield{author}{\bibinfo{person}{Julian Ashwin}, \bibinfo{person}{Aditya Chhabra}, {and} \bibinfo{person}{Vijayendra Rao}.} \bibinfo{year}{2023}\natexlab{}.
\newblock \showarticletitle{Using {Large} {Language} {Models} for {Qualitative} {Analysis} can {Introduce} {Serious} {Bias}}.
\newblock  (\bibinfo{year}{2023}).
\newblock
\urldef\tempurl%
\url{https://doi.org/10.48550/ARXIV.2309.17147}
\showDOI{\tempurl}


\bibitem[Bardzell and Bardzell(2016)]%
        {Bardzell:16}
\bibfield{author}{\bibinfo{person}{Jeffrey Bardzell} {and} \bibinfo{person}{Shaowen Bardzell}.} \bibinfo{year}{2016}\natexlab{}.
\newblock \showarticletitle{Humanistic {HCI}}.
\newblock \bibinfo{journal}{\emph{interactions}} \bibinfo{volume}{23}, \bibinfo{number}{2} (\bibinfo{date}{Feb.} \bibinfo{year}{2016}), \bibinfo{pages}{20--29}.
\newblock
\showISSN{1072-5520}
\urldef\tempurl%
\url{https://doi.org/10.1145/2888576}
\showDOI{\tempurl}


\bibitem[Bender et~al\mbox{.}(2021)]%
        {Bender:21}
\bibfield{author}{\bibinfo{person}{Emily~M. Bender}, \bibinfo{person}{Timnit Gebru}, \bibinfo{person}{Angelina McMillan-Major}, {and} \bibinfo{person}{Shmargaret Shmitchell}.} \bibinfo{year}{2021}\natexlab{}.
\newblock \showarticletitle{On the {Dangers} of {Stochastic} {Parrots}: {Can} {Language} {Models} {Be} {Too} {Big}?} \emph{(\bibinfo{series}{{FAccT} '21})}. \bibinfo{publisher}{Association for Computing Machinery}, \bibinfo{address}{New York, NY, USA}, \bibinfo{pages}{610--623}.
\newblock
\showISBNx{978-1-4503-8309-7}
\urldef\tempurl%
\url{https://doi.org/10.1145/3442188.3445922}
\showDOI{\tempurl}


\bibitem[Braun and Clarke(2019)]%
        {Braun:19}
\bibfield{author}{\bibinfo{person}{Virginia Braun} {and} \bibinfo{person}{Victoria Clarke}.} \bibinfo{year}{2019}\natexlab{}.
\newblock \showarticletitle{Reflecting on reflexive thematic analysis}.
\newblock \bibinfo{journal}{\emph{Qualitative Research in Sport, Exercise and Health}} \bibinfo{volume}{11}, \bibinfo{number}{4} (\bibinfo{year}{2019}), \bibinfo{pages}{589--597}.
\newblock
\urldef\tempurl%
\url{https://doi.org/10.1080/2159676X.2019.1628806}
\showDOI{\tempurl}


\bibitem[Brown et~al\mbox{.}(2020)]%
        {Brown:20}
\bibfield{author}{\bibinfo{person}{Tom Brown}, \bibinfo{person}{Benjamin Mann}, \bibinfo{person}{Nick Ryder}, \bibinfo{person}{Melanie Subbiah}, \bibinfo{person}{Jared~D Kaplan}, \bibinfo{person}{Prafulla Dhariwal}, \bibinfo{person}{Arvind Neelakantan}, \bibinfo{person}{Pranav Shyam}, \bibinfo{person}{Girish Sastry}, \bibinfo{person}{Amanda Askell}, \bibinfo{person}{Sandhini Agarwal}, \bibinfo{person}{Ariel Herbert-Voss}, \bibinfo{person}{Gretchen Krueger}, \bibinfo{person}{Tom Henighan}, \bibinfo{person}{Rewon Child}, \bibinfo{person}{Aditya Ramesh}, \bibinfo{person}{Daniel Ziegler}, \bibinfo{person}{Jeffrey Wu}, \bibinfo{person}{Clemens Winter}, \bibinfo{person}{Chris Hesse}, \bibinfo{person}{Mark Chen}, \bibinfo{person}{Eric Sigler}, \bibinfo{person}{Mateusz Litwin}, \bibinfo{person}{Scott Gray}, \bibinfo{person}{Benjamin Chess}, \bibinfo{person}{Jack Clark}, \bibinfo{person}{Christopher Berner}, \bibinfo{person}{Sam McCandlish}, \bibinfo{person}{Alec Radford}, \bibinfo{person}{Ilya Sutskever}, {and}
  \bibinfo{person}{Dario Amodei}.} \bibinfo{year}{2020}\natexlab{}.
\newblock \showarticletitle{Language Models are Few-Shot Learners}. In \bibinfo{booktitle}{\emph{Advances in Neural Information Processing Systems}}, \bibfield{editor}{\bibinfo{person}{H.~Larochelle}, \bibinfo{person}{M.~Ranzato}, \bibinfo{person}{R.~Hadsell}, \bibinfo{person}{M.F. Balcan}, {and} \bibinfo{person}{H.~Lin}} (Eds.), Vol.~\bibinfo{volume}{33}. \bibinfo{publisher}{Curran Associates, Inc.}, \bibinfo{pages}{1877--1901}.
\newblock
\urldef\tempurl%
\url{https://proceedings.neurips.cc/paper_files/paper/2020/file/1457c0d6bfcb4967418bfb8ac142f64a-Paper.pdf}
\showURL{%
\tempurl}


\bibitem[Cross(2011)]%
        {Cross:11}
\bibfield{author}{\bibinfo{person}{Nigel Cross}.} \bibinfo{year}{2011}\natexlab{}.
\newblock \bibinfo{booktitle}{\emph{Design Thinking: Understanding How Designers Think and Work}}.
\newblock \bibinfo{publisher}{Berg}, \bibinfo{address}{Oxford}.
\newblock
\showISBNx{9781847886361}
\urldef\tempurl%
\url{https://doi.org/10.5040/9781474293884}
\showDOI{\tempurl}


\bibitem[Dai et~al\mbox{.}(2023)]%
        {Dai:23}
\bibfield{author}{\bibinfo{person}{Shih-Chieh Dai}, \bibinfo{person}{Aiping Xiong}, {and} \bibinfo{person}{Lun-Wei Ku}.} \bibinfo{year}{2023}\natexlab{}.
\newblock \showarticletitle{{LLM}-in-the-loop: Leveraging Large Language Model for Thematic Analysis}. In \bibinfo{booktitle}{\emph{Findings of the Association for Computational Linguistics: EMNLP 2023}}, \bibfield{editor}{\bibinfo{person}{Houda Bouamor}, \bibinfo{person}{Juan Pino}, {and} \bibinfo{person}{Kalika Bali}} (Eds.). \bibinfo{publisher}{Association for Computational Linguistics}, \bibinfo{address}{Singapore}, \bibinfo{pages}{9993--10001}.
\newblock
\urldef\tempurl%
\url{https://doi.org/10.18653/v1/2023.findings-emnlp.669}
\showDOI{\tempurl}


\bibitem[Dietterich(1995)]%
        {dietterich1995overfitting}
\bibfield{author}{\bibinfo{person}{Tom Dietterich}.} \bibinfo{year}{1995}\natexlab{}.
\newblock \showarticletitle{Overfitting and undercomputing in machine learning}.
\newblock \bibinfo{journal}{\emph{ACM computing surveys (CSUR)}} \bibinfo{volume}{27}, \bibinfo{number}{3} (\bibinfo{year}{1995}), \bibinfo{pages}{326--327}.
\newblock


\bibitem[{Dovetail}({[n.\,d.]})]%
        {Dovetail:24}
\bibfield{author}{\bibinfo{person}{{Dovetail}}.} \bibinfo{year}{[n.\,d.]}\natexlab{}.
\newblock \bibinfo{title}{Customer Insights Hub}.
\newblock \bibinfo{howpublished}{\url{https://dovetail.com/}}.
\newblock
\newblock
\shownote{Accessed: 2024-12-31}.


\bibitem[Dubuisson and Jain(1994)]%
        {dubuisson1994modified}
\bibfield{author}{\bibinfo{person}{M-P Dubuisson} {and} \bibinfo{person}{Anil~K Jain}.} \bibinfo{year}{1994}\natexlab{}.
\newblock \showarticletitle{A modified Hausdorff distance for object matching}. In \bibinfo{booktitle}{\emph{Proceedings of 12th international conference on pattern recognition}}, Vol.~\bibinfo{volume}{1}. IEEE, \bibinfo{pages}{566--568}.
\newblock


\bibitem[Dunivin(2024)]%
        {Dunvin:24}
\bibfield{author}{\bibinfo{person}{Zackary~Okun Dunivin}.} \bibinfo{year}{2024}\natexlab{}.
\newblock \showarticletitle{Scalable {Qualitative} {Coding} with {LLMs}: {Chain}-of-{Thought} {Reasoning} {Matches} {Human} {Performance} in {Some} {Hermeneutic} {Tasks}}.
\newblock  (\bibinfo{year}{2024}).
\newblock
\urldef\tempurl%
\url{https://doi.org/10.48550/ARXIV.2401.15170}
\showDOI{\tempurl}


\bibitem[Dunne and Raby(2013)]%
        {Dunne:13}
\bibfield{author}{\bibinfo{person}{Anthony Dunne} {and} \bibinfo{person}{Fiona Raby}.} \bibinfo{year}{2013}\natexlab{}.
\newblock \bibinfo{booktitle}{\emph{Speculative {Everything}: {Design}, {Fiction}, and {Social} {Dreaming}}}.
\newblock \bibinfo{publisher}{The MIT Press}.
\newblock
\showISBNx{978-0-262-01984-2}


\bibitem[Feuston and Brubaker(2021)]%
        {Feuston:21}
\bibfield{author}{\bibinfo{person}{Jessica~L. Feuston} {and} \bibinfo{person}{Jed~R. Brubaker}.} \bibinfo{year}{2021}\natexlab{}.
\newblock \showarticletitle{Putting Tools in Their Place: The Role of Time and Perspective in Human-AI Collaboration for Qualitative Analysis}.
\newblock \bibinfo{journal}{\emph{Proc. ACM Hum.-Comput. Interact.}} \bibinfo{volume}{5}, \bibinfo{number}{CSCW2}, Article \bibinfo{articleno}{469} (\bibinfo{year}{2021}), \bibinfo{numpages}{25}~pages.
\newblock
\urldef\tempurl%
\url{https://doi.org/10.1145/3479856}
\showDOI{\tempurl}


\bibitem[Gebreegziabher et~al\mbox{.}(2023)]%
        {Gebreegziabher:23}
\bibfield{author}{\bibinfo{person}{Simret~Araya Gebreegziabher}, \bibinfo{person}{Zheng Zhang}, \bibinfo{person}{Xiaohang Tang}, \bibinfo{person}{Yihao Meng}, \bibinfo{person}{Elena~L. Glassman}, {and} \bibinfo{person}{Toby Jia-Jun Li}.} \bibinfo{year}{2023}\natexlab{}.
\newblock \showarticletitle{PaTAT: Human-AI Collaborative Qualitative Coding with Explainable Interactive Rule Synthesis}. In \bibinfo{booktitle}{\emph{Proceedings of the 2023 CHI Conference on Human Factors in Computing Systems}} (Hamburg, Germany) \emph{(\bibinfo{series}{CHI '23})}. \bibinfo{publisher}{Association for Computing Machinery}, \bibinfo{address}{New York, NY, USA}, Article \bibinfo{articleno}{362}, \bibinfo{numpages}{19}~pages.
\newblock
\showISBNx{9781450394215}
\urldef\tempurl%
\url{https://doi.org/10.1145/3544548.3581352}
\showDOI{\tempurl}


\bibitem[Gray and Malins(2004)]%
        {Gray:04}
\bibfield{author}{\bibinfo{person}{Carole Gray} {and} \bibinfo{person}{Julian Malins}.} \bibinfo{year}{2004}\natexlab{}.
\newblock \bibinfo{booktitle}{\emph{Visualizing Research: A Guide to the Research Process in Art and Design}}.
\newblock \bibinfo{publisher}{Routledge}, \bibinfo{address}{London}.
\newblock
\showISBNx{9780754635772}


\bibitem[H\"{a}m\"{a}l\"{a}inen et~al\mbox{.}(2023)]%
        {Hamalainen:23}
\bibfield{author}{\bibinfo{person}{Perttu H\"{a}m\"{a}l\"{a}inen}, \bibinfo{person}{Mikke Tavast}, {and} \bibinfo{person}{Anton Kunnari}.} \bibinfo{year}{2023}\natexlab{}.
\newblock \showarticletitle{Evaluating Large Language Models in Generating Synthetic HCI Research Data: A Case Study}. In \bibinfo{booktitle}{\emph{Proceedings of the 2023 CHI Conference on Human Factors in Computing Systems}} (Hamburg, Germany) \emph{(\bibinfo{series}{CHI '23})}. \bibinfo{publisher}{Association for Computing Machinery}, \bibinfo{address}{New York, NY, USA}, Article \bibinfo{articleno}{433}, \bibinfo{numpages}{19}~pages.
\newblock
\showISBNx{9781450394215}
\urldef\tempurl%
\url{https://doi.org/10.1145/3544548.3580688}
\showDOI{\tempurl}


\bibitem[Hamilton et~al\mbox{.}(2023)]%
        {Hamilton:23}
\bibfield{author}{\bibinfo{person}{Leah Hamilton}, \bibinfo{person}{Desha Elliott}, \bibinfo{person}{Aaron Quick}, \bibinfo{person}{Simone Smith}, {and} \bibinfo{person}{Victoria Choplin}.} \bibinfo{year}{2023}\natexlab{}.
\newblock \showarticletitle{Exploring the {Use} of {AI} in {Qualitative} {Analysis}: {A} {Comparative} {Study} of {Guaranteed} {Income} {Data}}.
\newblock \bibinfo{journal}{\emph{International Journal of Qualitative Methods}}  \bibinfo{volume}{22} (\bibinfo{date}{Oct.} \bibinfo{year}{2023}), \bibinfo{pages}{16094069231201504}.
\newblock
\showISSN{1609-4069}
\urldef\tempurl%
\url{https://doi.org/10.1177/16094069231201504}
\showDOI{\tempurl}


\bibitem[Hong et~al\mbox{.}(2022)]%
        {Hong:22}
\bibfield{author}{\bibinfo{person}{Matt-Heun Hong}, \bibinfo{person}{Lauren~A. Marsh}, \bibinfo{person}{Jessica~L. Feuston}, \bibinfo{person}{Janet Ruppert}, \bibinfo{person}{Jed~R. Brubaker}, {and} \bibinfo{person}{Danielle~Albers Szafir}.} \bibinfo{year}{2022}\natexlab{}.
\newblock \showarticletitle{Scholastic: Graphical Human-AI Collaboration for Inductive and Interpretive Text Analysis}. In \bibinfo{booktitle}{\emph{Proceedings of the 35th Annual ACM Symposium on User Interface Software and Technology}} (Bend, OR, USA) \emph{(\bibinfo{series}{UIST '22})}. \bibinfo{publisher}{Association for Computing Machinery}, \bibinfo{address}{New York, NY, USA}, Article \bibinfo{articleno}{30}, \bibinfo{numpages}{12}~pages.
\newblock
\showISBNx{9781450393201}
\urldef\tempurl%
\url{https://doi.org/10.1145/3526113.3545681}
\showDOI{\tempurl}


\bibitem[Jacovi et~al\mbox{.}(2021)]%
        {jacovi:21}
\bibfield{author}{\bibinfo{person}{Alon Jacovi}, \bibinfo{person}{Ana Marasovi\'{c}}, \bibinfo{person}{Tim Miller}, {and} \bibinfo{person}{Yoav Goldberg}.} \bibinfo{year}{2021}\natexlab{}.
\newblock \showarticletitle{Formalizing Trust in Artificial Intelligence: Prerequisites, Causes and Goals of Human Trust in AI}. In \bibinfo{booktitle}{\emph{Proceedings of the 2021 ACM Conference on Fairness, Accountability, and Transparency}} (Virtual Event, Canada) \emph{(\bibinfo{series}{FAccT '21})}. \bibinfo{publisher}{Association for Computing Machinery}, \bibinfo{address}{New York, NY, USA}, \bibinfo{pages}{624–635}.
\newblock
\showISBNx{9781450383097}
\urldef\tempurl%
\url{https://doi.org/10.1145/3442188.3445923}
\showDOI{\tempurl}


\bibitem[Jiang et~al\mbox{.}(2021)]%
        {Jiang:21}
\bibfield{author}{\bibinfo{person}{Jialun~Aaron Jiang}, \bibinfo{person}{Kandrea Wade}, \bibinfo{person}{Casey Fiesler}, {and} \bibinfo{person}{Jed~R. Brubaker}.} \bibinfo{year}{2021}\natexlab{}.
\newblock \showarticletitle{Supporting Serendipity: Opportunities and Challenges for Human-AI Collaboration in Qualitative Analysis}.
\newblock \bibinfo{journal}{\emph{Proc. ACM Hum.-Comput. Interact.}} \bibinfo{volume}{5}, \bibinfo{number}{CSCW1}, Article \bibinfo{articleno}{94} (\bibinfo{year}{2021}), \bibinfo{numpages}{23}~pages.
\newblock
\urldef\tempurl%
\url{https://doi.org/10.1145/3449168}
\showDOI{\tempurl}


\bibitem[Kluyver et~al\mbox{.}(2016)]%
        {Kluyver:16}
\bibfield{author}{\bibinfo{person}{Thomas Kluyver}, \bibinfo{person}{Benjamin Ragan-Kelley}, \bibinfo{person}{Fernando P{\'e}rez}, \bibinfo{person}{Brian Granger}, \bibinfo{person}{Matthias Bussonnier}, \bibinfo{person}{Jonathan Frederic}, \bibinfo{person}{Kyle Kelley}, \bibinfo{person}{Jessica Hamrick}, \bibinfo{person}{Jason Grout}, \bibinfo{person}{Sylvain Corlay}, \bibinfo{person}{Paul Ivanov}, \bibinfo{person}{Dami{\'a}n Avila}, \bibinfo{person}{Safia Abdalla}, \bibinfo{person}{Carol Willing}, {and} \bibinfo{person}{Jupyter development team}.} \bibinfo{year}{2016}\natexlab{}.
\newblock \showarticletitle{Jupyter Notebooks - a publishing format for reproducible computational workflows}. In \bibinfo{booktitle}{\emph{Positioning and Power in Academic Publishing: Players, Agents and Agendas}}, \bibfield{editor}{\bibinfo{person}{Fernando Loizides} {and} \bibinfo{person}{Birgit Scmidt}} (Eds.). \bibinfo{publisher}{IOS Press}, \bibinfo{address}{Netherlands}, \bibinfo{pages}{87--90}.
\newblock
\urldef\tempurl%
\url{https://eprints.soton.ac.uk/403913/}
\showURL{%
\tempurl}


\bibitem[Krippendorff(2005)]%
        {Krippendorff:05}
\bibfield{author}{\bibinfo{person}{Klaus Krippendorff}.} \bibinfo{year}{2005}\natexlab{}.
\newblock \bibinfo{booktitle}{\emph{The Semantic Turn: A New Foundation for Design} (\bibinfo{edition}{1.} ed.)}.
\newblock \bibinfo{publisher}{CRC Press}, \bibinfo{address}{Boca Raton, FL, USA}.
\newblock
\showISBNx{9780429219122}
\urldef\tempurl%
\url{https://doi.org/10.4324/9780203299951}
\showDOI{\tempurl}


\bibitem[Lee and See(2004)]%
        {lee:04}
\bibfield{author}{\bibinfo{person}{John~D. Lee} {and} \bibinfo{person}{Katrina~A. See}.} \bibinfo{year}{2004}\natexlab{}.
\newblock \showarticletitle{Trust in Automation: Designing for Appropriate Reliance}.
\newblock \bibinfo{journal}{\emph{Human Factors}} \bibinfo{volume}{46}, \bibinfo{number}{1} (\bibinfo{year}{2004}), \bibinfo{pages}{50--80}.
\newblock
\urldef\tempurl%
\url{https://doi.org/10.1518/hfes.46.1.50\_30392}
\showDOI{\tempurl}


\bibitem[Lopez-Fierro and Nguyen(2024)]%
        {Lopez:24}
\bibfield{author}{\bibinfo{person}{Sar{\'\i}ah Lopez-Fierro} {and} \bibinfo{person}{Ha Nguyen}.} \bibinfo{year}{2024}\natexlab{}.
\newblock \showarticletitle{Making Human-AI Contributions Transparent in Qualitative Coding}. In \bibinfo{booktitle}{\emph{Proceedings of the 17th International Conference on Computer-Supported Collaborative Learning-CSCL 2024, pp. 3-10}}. International Society of the Learning Sciences.
\newblock


\bibitem[Lucero(2012)]%
        {Lucero18}
\bibfield{author}{\bibinfo{person}{Andr\'{e}s Lucero}.} \bibinfo{year}{2012}\natexlab{}.
\newblock \showarticletitle{Framing, aligning, paradoxing, abstracting, and directing: how design mood boards work}. In \bibinfo{booktitle}{\emph{Proceedings of the Designing Interactive Systems Conference}} (Newcastle Upon Tyne, United Kingdom) \emph{(\bibinfo{series}{DIS '12})}. \bibinfo{publisher}{Association for Computing Machinery}, \bibinfo{address}{New York, NY, USA}, \bibinfo{pages}{438–447}.
\newblock
\showISBNx{9781450312103}
\urldef\tempurl%
\url{https://doi.org/10.1145/2317956.2318021}
\showDOI{\tempurl}


\bibitem[Lucero(2015)]%
        {Lucero15}
\bibfield{author}{\bibinfo{person}{Andr{\'e}s Lucero}.} \bibinfo{year}{2015}\natexlab{}.
\newblock \showarticletitle{Using Affinity Diagrams to Evaluate Interactive Prototypes}. In \bibinfo{booktitle}{\emph{Human-Computer Interaction -- INTERACT 2015}}, \bibfield{editor}{\bibinfo{person}{Julio Abascal}, \bibinfo{person}{Simone Barbosa}, \bibinfo{person}{Mirko Fetter}, \bibinfo{person}{Tom Gross}, \bibinfo{person}{Philippe Palanque}, {and} \bibinfo{person}{Marco Winckler}} (Eds.). \bibinfo{publisher}{Springer International Publishing}, \bibinfo{address}{Cham}, \bibinfo{pages}{231--248}.
\newblock
\showISBNx{978-3-319-22668-2}


\bibitem[Marathe and Toyama(2018)]%
        {Marathe:18}
\bibfield{author}{\bibinfo{person}{Megh Marathe} {and} \bibinfo{person}{Kentaro Toyama}.} \bibinfo{year}{2018}\natexlab{}.
\newblock \showarticletitle{Semi-Automated Coding for Qualitative Research: A User-Centered Inquiry and Initial Prototypes}. In \bibinfo{booktitle}{\emph{Proceedings of the 2018 CHI Conference on Human Factors in Computing Systems}} (Montreal QC, Canada) \emph{(\bibinfo{series}{CHI '18})}. \bibinfo{publisher}{Association for Computing Machinery}, \bibinfo{address}{New York, NY, USA}, \bibinfo{pages}{1–12}.
\newblock
\showISBNx{9781450356206}
\urldef\tempurl%
\url{https://doi.org/10.1145/3173574.3173922}
\showDOI{\tempurl}


\bibitem[Mayer et~al\mbox{.}(1995)]%
        {mayer:95}
\bibfield{author}{\bibinfo{person}{Roger~C. Mayer}, \bibinfo{person}{James~H. Davis}, {and} \bibinfo{person}{F.~David Schoorman}.} \bibinfo{year}{1995}\natexlab{}.
\newblock \showarticletitle{An Integrative Model of Organizational Trust}.
\newblock \bibinfo{journal}{\emph{The Academy of Management Review}} \bibinfo{volume}{20}, \bibinfo{number}{3} (\bibinfo{year}{1995}), \bibinfo{pages}{709--734}.
\newblock
\showISSN{03637425}
\urldef\tempurl%
\url{https://doi.org/10.2307/258792}
\showDOI{\tempurl}


\bibitem[Micheli et~al\mbox{.}(2019)]%
        {Micheli:19}
\bibfield{author}{\bibinfo{person}{Pietro Micheli}, \bibinfo{person}{Sarah J.~S. Wilner}, \bibinfo{person}{Sabeen~Hussain Bhatti}, \bibinfo{person}{Matteo Mura}, {and} \bibinfo{person}{Michael~B. Beverland}.} \bibinfo{year}{2019}\natexlab{}.
\newblock \showarticletitle{Doing {Design} {Thinking}: {Conceptual} {Review}, {Synthesis}, and {Research} {Agenda}}.
\newblock \bibinfo{journal}{\emph{Journal of Product Innovation Management}} \bibinfo{volume}{36}, \bibinfo{number}{2} (\bibinfo{year}{2019}), \bibinfo{pages}{124--148}.
\newblock
\showISSN{1540-5885}
\urldef\tempurl%
\url{https://doi.org/10.1111/jpim.12466}
\showDOI{\tempurl}


\bibitem[Miles et~al\mbox{.}(2014)]%
        {Miles:14}
\bibfield{author}{\bibinfo{person}{Matthew Miles}, \bibinfo{person}{A.~Michael Huberman}, {and} \bibinfo{person}{Johnny Saldaña}.} \bibinfo{year}{2014}\natexlab{}.
\newblock \bibinfo{booktitle}{\emph{Qualitative Data Analysis: A Methods Sourcebook} (\bibinfo{edition}{3.} ed.)}.
\newblock \bibinfo{publisher}{SAGE Publications, Inc}.
\newblock
\showISBNx{978-1-4522-5787-7}


\bibitem[Muratovski et~al\mbox{.}(2022)]%
        {Muratovski:22}
\bibfield{author}{\bibinfo{person}{Gjoko Muratovski}, \bibinfo{person}{Ken Friedman}, \bibinfo{person}{Don Norman}, {and} \bibinfo{person}{Steven Heller}.} \bibinfo{year}{2022}\natexlab{}.
\newblock \bibinfo{booktitle}{\emph{Research for Designers: A Guide to Methods and Practice} (\bibinfo{edition}{2.} ed.)}.
\newblock \bibinfo{publisher}{SAGE Publications}, \bibinfo{address}{London}.
\newblock
\showISBNx{978-1-5297-0816-5}


\bibitem[Oksanen(2024)]%
        {Oksanen:24}
\bibfield{author}{\bibinfo{person}{Joel Oksanen}.} \bibinfo{year}{2024}\natexlab{}.
\newblock \showarticletitle{Bridging the {Integrity} {Gap}: {Towards} {AI}-assisted {Design} {Research}}.
\newblock \bibinfo{journal}{\emph{Extended Abstracts of the 2024 CHI Conference on Human Factors in Computing Systems}} (\bibinfo{date}{May} \bibinfo{year}{2024}), \bibinfo{pages}{1--5}.
\newblock
\showISSN{9798400703317}
\urldef\tempurl%
\url{https://doi.org/10.1145/3613905.3647962}
\showDOI{\tempurl}


\bibitem[OpenAI(2024a)]%
        {OpenAI:24:2}
\bibfield{author}{\bibinfo{person}{OpenAI}.} \bibinfo{year}{2024}\natexlab{a}.
\newblock \bibinfo{title}{New embedding models and API updates}.
\newblock \bibinfo{howpublished}{https://openai.com/index/new-embedding-models-and-api-updates/}.
\newblock
\newblock
\shownote{Accessed: 2025-01-10}.


\bibitem[OpenAI(2024b)]%
        {OpenAI:24:3}
\bibfield{author}{\bibinfo{person}{OpenAI}.} \bibinfo{year}{2024}\natexlab{b}.
\newblock \bibinfo{title}{What are tokens and how to count them?}
\newblock \bibinfo{howpublished}{https://help.openai.com/en/articles/4936856-what-are-tokens-and-how-to-count-them}.
\newblock
\newblock
\shownote{Accessed: 2025-01-10}.


\bibitem[OpenAI(nd)]%
        {OpenAI:24}
\bibfield{author}{\bibinfo{person}{OpenAI}.} \bibinfo{year}{n.d.}\natexlab{}.
\newblock \bibinfo{title}{Models - OpenAI API}.
\newblock \bibinfo{howpublished}{\url{https://platform.openai.com/docs/models/}}.
\newblock
\newblock
\shownote{Accessed: 2025-01-08}.


\bibitem[Real and Vargas(1996)]%
        {real1996probabilistic}
\bibfield{author}{\bibinfo{person}{Raimundo Real} {and} \bibinfo{person}{Juan~M Vargas}.} \bibinfo{year}{1996}\natexlab{}.
\newblock \showarticletitle{The probabilistic basis of Jaccard's index of similarity}.
\newblock \bibinfo{journal}{\emph{Systematic biology}} \bibinfo{volume}{45}, \bibinfo{number}{3} (\bibinfo{year}{1996}), \bibinfo{pages}{380--385}.
\newblock


\bibitem[Rezatofighi et~al\mbox{.}(2019)]%
        {rezatofighi2019generalized}
\bibfield{author}{\bibinfo{person}{Hamid Rezatofighi}, \bibinfo{person}{Nathan Tsoi}, \bibinfo{person}{JunYoung Gwak}, \bibinfo{person}{Amir Sadeghian}, \bibinfo{person}{Ian Reid}, {and} \bibinfo{person}{Silvio Savarese}.} \bibinfo{year}{2019}\natexlab{}.
\newblock \showarticletitle{Generalized intersection over union: A metric and a loss for bounding box regression}. In \bibinfo{booktitle}{\emph{Proceedings of the IEEE/CVF conference on computer vision and pattern recognition}}. \bibinfo{pages}{658--666}.
\newblock


\bibitem[Sanders and Stappers(2008)]%
        {Sanders:08}
\bibfield{author}{\bibinfo{person}{Elizabeth B.-N. Sanders} {and} \bibinfo{person}{Pieter~Jan Stappers}.} \bibinfo{year}{2008}\natexlab{}.
\newblock \showarticletitle{Co-creation and the new landscapes of design}.
\newblock \bibinfo{journal}{\emph{CoDesign}} \bibinfo{volume}{4}, \bibinfo{number}{1} (\bibinfo{date}{March} \bibinfo{year}{2008}), \bibinfo{pages}{5--18}.
\newblock
\showISSN{1571-0882}
\urldef\tempurl%
\url{https://doi.org/10.1080/15710880701875068}
\showDOI{\tempurl}


\bibitem[Sch{\"o}n(1983)]%
        {Schon:83}
\bibfield{author}{\bibinfo{person}{Donald~A. Sch{\"o}n}.} \bibinfo{year}{1983}\natexlab{}.
\newblock \bibinfo{booktitle}{\emph{The Reflective Practitioner: How Professionals Think in Action}}.
\newblock \bibinfo{publisher}{Basic Books}, \bibinfo{address}{New York, NY}.
\newblock
\showISBNx{978-0465068784}


\bibitem[Sinha et~al\mbox{.}(2024)]%
        {Sinha:24}
\bibfield{author}{\bibinfo{person}{Ravi Sinha}, \bibinfo{person}{Idris Solola}, \bibinfo{person}{Ha Nguyen}, \bibinfo{person}{Hillary Swanson}, {and} \bibinfo{person}{LuEttaMae Lawrence}.} \bibinfo{year}{2024}\natexlab{}.
\newblock \showarticletitle{The {Role} of {Generative} {AI} in {Qualitative} {Research}: {GPT}-4's {Contributions} to a {Grounded} {Theory} {Analysis}} \emph{(\bibinfo{series}{{LDT} '24})}. \bibinfo{publisher}{Association for Computing Machinery}, \bibinfo{address}{New York, NY, USA}, \bibinfo{pages}{17--25}.
\newblock
\showISBNx{9798400717222}
\urldef\tempurl%
\url{https://doi.org/10.1145/3663433.3663456}
\showDOI{\tempurl}


\bibitem[Touvron et~al\mbox{.}(2023)]%
        {Touvron:23}
\bibfield{author}{\bibinfo{person}{Hugo Touvron}, \bibinfo{person}{Louis Martin}, \bibinfo{person}{Kevin Stone}, \bibinfo{person}{Peter Albert}, \bibinfo{person}{Amjad Almahairi}, \bibinfo{person}{Yasmine Babaei}, \bibinfo{person}{Nikolay Bashlykov}, \bibinfo{person}{Soumya Batra}, \bibinfo{person}{Prajjwal Bhargava}, \bibinfo{person}{Shruti Bhosale}, \bibinfo{person}{Dan Bikel}, \bibinfo{person}{Lukas Blecher}, \bibinfo{person}{Cristian~Canton Ferrer}, \bibinfo{person}{Moya Chen}, \bibinfo{person}{Guillem Cucurull}, \bibinfo{person}{David Esiobu}, \bibinfo{person}{Jude Fernandes}, \bibinfo{person}{Jeremy Fu}, \bibinfo{person}{Wenyin Fu}, \bibinfo{person}{Brian Fuller}, \bibinfo{person}{Cynthia Gao}, \bibinfo{person}{Vedanuj Goswami}, \bibinfo{person}{Naman Goyal}, \bibinfo{person}{Anthony Hartshorn}, \bibinfo{person}{Saghar Hosseini}, \bibinfo{person}{Rui Hou}, \bibinfo{person}{Hakan Inan}, \bibinfo{person}{Marcin Kardas}, \bibinfo{person}{Viktor Kerkez}, \bibinfo{person}{Madian Khabsa},
  \bibinfo{person}{Isabel Kloumann}, \bibinfo{person}{Artem Korenev}, \bibinfo{person}{Punit~Singh Koura}, \bibinfo{person}{Marie-Anne Lachaux}, \bibinfo{person}{Thibaut Lavril}, \bibinfo{person}{Jenya Lee}, \bibinfo{person}{Diana Liskovich}, \bibinfo{person}{Yinghai Lu}, \bibinfo{person}{Yuning Mao}, \bibinfo{person}{Xavier Martinet}, \bibinfo{person}{Todor Mihaylov}, \bibinfo{person}{Pushkar Mishra}, \bibinfo{person}{Igor Molybog}, \bibinfo{person}{Yixin Nie}, \bibinfo{person}{Andrew Poulton}, \bibinfo{person}{Jeremy Reizenstein}, \bibinfo{person}{Rashi Rungta}, \bibinfo{person}{Kalyan Saladi}, \bibinfo{person}{Alan Schelten}, \bibinfo{person}{Ruan Silva}, \bibinfo{person}{Eric~Michael Smith}, \bibinfo{person}{Ranjan Subramanian}, \bibinfo{person}{Xiaoqing~Ellen Tan}, \bibinfo{person}{Binh Tang}, \bibinfo{person}{Ross Taylor}, \bibinfo{person}{Adina Williams}, \bibinfo{person}{Jian~Xiang Kuan}, \bibinfo{person}{Puxin Xu}, \bibinfo{person}{Zheng Yan}, \bibinfo{person}{Iliyan Zarov}, \bibinfo{person}{Yuchen
  Zhang}, \bibinfo{person}{Angela Fan}, \bibinfo{person}{Melanie Kambadur}, \bibinfo{person}{Sharan Narang}, \bibinfo{person}{Aurelien Rodriguez}, \bibinfo{person}{Robert Stojnic}, \bibinfo{person}{Sergey Edunov}, {and} \bibinfo{person}{Thomas Scialom}.} \bibinfo{year}{2023}\natexlab{}.
\newblock \bibinfo{title}{Llama 2: Open Foundation and Fine-Tuned Chat Models}.
\newblock
\newblock
\showeprint[arxiv]{2307.09288}~[cs.CL]


\bibitem[Verma and Nidhi(2017)]%
        {verma2017extractive}
\bibfield{author}{\bibinfo{person}{Sukriti Verma} {and} \bibinfo{person}{Vagisha Nidhi}.} \bibinfo{year}{2017}\natexlab{}.
\newblock \showarticletitle{Extractive summarization using deep learning}.
\newblock \bibinfo{journal}{\emph{arXiv preprint arXiv:1708.04439}} (\bibinfo{year}{2017}).
\newblock


\bibitem[Vianna et~al\mbox{.}(2024)]%
        {Vianna:24}
\bibfield{author}{\bibinfo{person}{Leonardo~Silva Vianna}, \bibinfo{person}{Alexandre~Leopoldo Gonçalves}, {and} \bibinfo{person}{João~Artur Souza}.} \bibinfo{year}{2024}\natexlab{}.
\newblock \showarticletitle{Analysis of learning curves in predictive modeling using exponential curve fitting with an asymptotic approach}.
\newblock \bibinfo{journal}{\emph{PLOS ONE}} \bibinfo{volume}{19}, \bibinfo{number}{4} (\bibinfo{date}{04} \bibinfo{year}{2024}), \bibinfo{pages}{1--23}.
\newblock
\urldef\tempurl%
\url{https://doi.org/10.1371/journal.pone.0299811}
\showDOI{\tempurl}


\bibitem[Violeta~Clemente and Pombo(2017)]%
        {Clemente:17}
\bibfield{author}{\bibinfo{person}{Katja~Tschimmel Violeta~Clemente} {and} \bibinfo{person}{Fátima Pombo}.} \bibinfo{year}{2017}\natexlab{}.
\newblock \showarticletitle{A Future Scenario for a Methodological Approach applied to PhD Design Research. Development of an Analytical Canvas}.
\newblock \bibinfo{journal}{\emph{The Design Journal}} \bibinfo{volume}{20}, \bibinfo{number}{sup1} (\bibinfo{year}{2017}), \bibinfo{pages}{S792--S802}.
\newblock
\urldef\tempurl%
\url{https://doi.org/10.1080/14606925.2017.1353025}
\showDOI{\tempurl}


\bibitem[Xiao et~al\mbox{.}(2023)]%
        {Xiao:23}
\bibfield{author}{\bibinfo{person}{Ziang Xiao}, \bibinfo{person}{Xingdi Yuan}, \bibinfo{person}{Q.~Vera Liao}, \bibinfo{person}{Rania Abdelghani}, {and} \bibinfo{person}{Pierre-Yves Oudeyer}.} \bibinfo{year}{2023}\natexlab{}.
\newblock \showarticletitle{Supporting Qualitative Analysis with Large Language Models: Combining Codebook with GPT-3 for Deductive Coding}. In \bibinfo{booktitle}{\emph{Companion Proceedings of the 28th International Conference on Intelligent User Interfaces}} (Sydney, NSW, Australia) \emph{(\bibinfo{series}{IUI '23 Companion})}. \bibinfo{publisher}{Association for Computing Machinery}, \bibinfo{address}{New York, NY, USA}, \bibinfo{pages}{75–78}.
\newblock
\showISBNx{9798400701078}
\urldef\tempurl%
\url{https://doi.org/10.1145/3581754.3584136}
\showDOI{\tempurl}


\bibitem[Zhao et~al\mbox{.}(2024)]%
        {Zhao:24}
\bibfield{author}{\bibinfo{person}{Fengxiang Zhao}, \bibinfo{person}{Fan Yu}, {and} \bibinfo{person}{Yi Shang}.} \bibinfo{year}{2024}\natexlab{}.
\newblock \showarticletitle{A {New} {Method} {Supporting} {Qualitative} {Data} {Analysis} {Through} {Prompt} {Generation} for {Inductive} {Coding}}. In \bibinfo{booktitle}{\emph{2024 {IEEE} {International} {Conference} on {Information} {Reuse} and {Integration} for {Data} {Science} ({IRI})}}. \bibinfo{pages}{164--169}.
\newblock
\urldef\tempurl%
\url{https://doi.org/10.1109/IRI62200.2024.00043}
\showDOI{\tempurl}
\newblock
\shownote{ISSN: 2835-5776}.


\end{thebibliography}

\appendix

\section{LLMCode Inductive Coding Prompt}
\label{appendix:prompt}

\doublebox{
  \begin{minipage}{0.95\textwidth}

You are an expert qualitative researcher. You are given a text to code inductively. Please carry out the following task: \\
- Respond by repeating the original text, but highlighting the coded statements by surrounding the statements with double asterisks, as if they were bolded text in a Markdown document. \\
- Include the associated code(s) immediately after the statement, separated by a semicolon and enclosed in <sup></sup> tags, as if they were superscript text in a Markdown document. \\
- Preserve exact formatting of the original text. Do not correct typos or remove unnecessary spaces.

\vspace{1em}
  
    \shadowbox{
      \begin{minipage}{0.95\textwidth}
        \textbf{Custom user-defined instructions}, e.g. \\
        - Do not code interviewer questions.
      \end{minipage}
    }

\vspace{1em}

Some examples of codes in the format "\{code\}: \{description\}". Please create new codes when needed:
    
\vspace{1em}
    
    \shadowbox{
      \begin{minipage}{0.95\textwidth}
        \textbf{Code descriptions (automatically generated by a separate LLM prompt for each new code)}, e.g. \\
        - travel frequency: Describes instances where the participant shares how often they travel. \\
        ...
      \end{minipage}
    }
    
\vspace{1em}
    
    Below, I first give you examples of the output you should produce given an example input. After that, I give you the actual input to process. The input may come from a thread of texts, and any preceding texts are added as context (labelled CONTEXT). Your task is to code only the last text (labelled TEXT).

\vspace{1em}

\shadowbox{
      \begin{minipage}{0.95\textwidth}
        \textbf{Few-shot examples}, e.g. \\
        EXAMPLE INPUT: \\
        CONTEXT: How often do you travel? \\
        TEXT: I travel quite often, or at least maybe four times a year. \\
        EXAMPLE OUTPUT: **I travel quite often**<sup>travel frequency</sup>, or at least maybe four times a year. \\
        ...
      \end{minipage}
    }
    
\vspace{1em}

ACTUAL INPUT:

\vspace{1em}

\shadowbox{
      \begin{minipage}{0.95\textwidth}
      \textbf{A single input text}, e.g. \\
        CONTEXT: In your own words, can you describe how the loyalty program works? \\
        TEXT: I think it starts on basic actually, basic, silver, gold. It's this typical good, better, best. Maybe they could spice that up a bit, make some more remarkable names or different colours.
      \end{minipage}
    }

  \end{minipage}
}

\end{document}